\documentclass{aa}
\usepackage[varg]{txfonts}
\usepackage{natbib}
\usepackage{graphicx}
\usepackage{amsmath}
\usepackage{xcolor}
\usepackage[normalem]{ulem}
\bibpunct{(}{)}{;}{a}{}{,} 

\begin{document}
\title{Stellar evolution models with overshooting based on 3-equation non-local theories}
\subtitle{II. Main-sequence models of A- and B-type stars} 
\author{F.\ Ahlborn\inst{1}
\and F.\ Kupka\inst{2,3,4}
\and A.\ Weiss\inst{1}
\and M. Flaskamp\inst{1}}
\institute{Max-Planck-Institut f\"ur Astrophysik, Karl-Schwarzschild-Stra\ss e 1, 85748 Garching, Germany\\
email: \url{fahlborn@mpa-garching.mpg.de}
\and Dept. Applied Mathematics and Physics, Univ. of Applied Sciences, Technikum Wien, H\"ochst\"adtplatz 6, A-1200 Wien, Austria
\and Wolfgang-Pauli-Institute c/o Faculty of Mathematics, University of Vienna, Oskar-Morgenstern-Platz 1, A-1090 Wien, Austria
\and Max-Planck-Institut f\"ur Sonnensystemforschung, Justus-von-Liebig-Weg 3, 37077 G\"ottingen, Germany
}
\date{Received xxx /
Accepted xxx }
\abstract
{Convective overshoot mixing is a critical ingredient of stellar structure models, but is treated in most cases by ad~hoc extensions of the mixing-length theory for convection.  Advanced theories which are both more physical and numerically treatable are needed. }
{Convective flows in stellar interiors are highly turbulent. This poses a number of numerical challenges for the modelling of convection in stellar interiors. We include an effective turbulence model into a 1D stellar evolution code in order to treat non-local effects within the same theory.}
{We use a turbulent convection model which relies on the solution of second order moment equations. We implement this into a state of the art 1D stellar evolution code. To overcome a deficit in the original form of the model, we take the dissipation due to buoyancy waves in the overshooting zone into account.}
{We compute stellar models of intermediate mass main-sequence stars between 1.5 and 8~$M_\odot$. Overshoot mixing from the convective core and modified temperature gradients within and above it emerge naturally as a solution of the turbulent convection model equations.}
{For a given set of model parameters the overshooting extent determined from the turbulent convection model is comparable to other overshooting descriptions, the free parameters of which had been adjusted to match observations. The relative size of the mixed cores decreases with decreasing stellar mass without additional adjustments. We find that the dissipation by buoyancy waves constitutes a necessary and  relevant extension of the turbulent convection model in use.}

\titlerunning{Turbulent convection for stellar evolution}
\authorrunning{F.\ Ahlborn et al.}
\maketitle
\section{Introduction}
Internal transport processes substantially shape the structure of stars. This concerns the transport of energy as well as the transport of chemical elements and angular momentum. In intermediate and high mass main-sequence stars, the energy released by nuclear fusion in the centre is transported by convection, which is the transport of energy by means of fluid motion. In deep convective zones, convection determines the temperature gradient and dominates the chemical mixing processes due to its efficiency. Hence, convection is a crucial aspect of stellar structure and evolution models.

Disregarding its importance, convection remains one of the major uncertainties in stellar structure and evolution modelling. The most commonly used description of convection for stellar models is the so-called mixing length theory (MLT) \citep{biermann1932,boehm1958}. Despite its simplicity --- MLT is a local and time-independent theory --- it was used very successfully over many years to model stellar interiors. With ever improving observational facilities and methods, however, more and more deficits of MLT become apparent. One of the main problems of MLT concerns the treatment of convective boundaries. When ignoring compositional effects, the acceleration drops to zero at the so-called Schwarzschild-boundary, while the velocity generally does not. From a theoretical point, it has been shown that convective motions should pass the Schwarzschild boundary and penetrate into the stable layers \citep{roxburgh1978,roxburgh1992,zahn1991}. In stable layers, convective motions are braked and the material carried with the flow mixes with the surroundings. In the literature, different terms are used to describe the processes at convective boundaries. \cite{zahn1991} differentiates between thermally efficient and inefficient convection. Thermally efficient convection is able to modify the model temperature gradient and is therefore referred to as subadiabatic penetration. Thermally inefficient convection leaves the temperature gradient unchanged while still mixing chemical elements. \cite{zahn1991} refers to this as overshoot mixing. Finally, the notion of convective entrainment refers to the continuous ingestion of material at convective boundaries into a convective region \citep{turner1986}. This effect was found to represent the convective boundary mixing in 3D simulations of an oxygen burning shell \citep{meakin2007}.

In the MLT picture, the convective velocities drop to zero at the formal Schwarzschild boundary, preventing convective mixing beyond this point.  In a physical configuration, however, only the acceleration of fluid elements disappears while the velocity generally remains finite. To include the effects of convective overshooting in stellar models, additional descriptions have to be applied. Early attempts were proposed, for example by \cite{saslaw1965} and \cite{shaviv1973}. For a critical review of these theories, we refer to \cite{renzini1987}. The main effect of convective overshooting on the stellar structure can be mimicked by introducing additional mixing at convective boundaries during stellar evolution, which we will refer to as ad~hoc overshooting. To account for the required mixing, \cite{freytag1996} introduce an additional diffusion constant, which decreases exponentially as a function of the distance to the Schwarzschild boundary. This is also known as ``diffusive overshooting''. Although this approach is originally based on 2D simulations of envelopes in A-type main-sequence stars and DA-type white dwarfs which have thin convective zones subject to strong radiative losses, it is commonly applied to all convective boundaries in stellar evolution models.

When assuming instantaneous mixing, ad~hoc overshooting is commonly introduced by extending the chemical composition of the convective region by a certain fraction of the pressure scale-height into the formally stable region. This approach is also known as ``step overshooting''. Neither of these descriptions provides constraints on the temperature gradient in the overshooting region or on the extent of this region. 

Due to the high Reynolds numbers, convection is a highly turbulent process. The dynamic time-scales of the involved flows are many orders of magnitude shorter than the nuclear time-scales of stellar evolution in most evolutionary phases. This poses serious problems for numerical descriptions of convection. Full 3D hydrodynamic simulations are computationally expensive and can cover physical time spans on the order of years only. In contrast, stellar evolution calculations usually need to be computed over at least a couple of million years. At the same time, the computational costs need to be low. Therefore, the direct inclusion of 3D hydrodynamics into stellar evolution calculations is not feasible on present day computers. This means that the effects of convection need to be included into the stellar evolution models by some other means. One way to include the results of 3D simulations into 1D stellar evolution codes is the Reynolds-averaged-Navier-Stokes (RANS) analysis, as for example outlined by \cite{viallet2013} and \cite{arnett2015}. 

However, since 3D simulations suffer from the aforementioned computing limitations, in practice this method is essentially a variant of the Reynolds stress approach \citep{keller1924,chou1945,andre1976a,canuto1992} and more generally speaking, of turbulent convection models (TCM). The main idea of TCM is to construct higher order moment equations from the hydrodynamic equations and reduce the dimensionality of the problem by averaging over two spatial directions. These equations describe the dynamics of convection. One main difference of TCM compared to MLT is the occurrence of transport terms. These terms describe the transport of physical quantities, for example the turbulent kinetic energy, by means of convective flows, and naturally lead to the emergence of phenomena like convective overshooting without any ad~hoc description. As these transport terms connect different layers, they are often termed ``non-local''. Furthermore, TCM provide the convective flux which allows computing the temperature gradient also in the overshooting region. To date, a number of TCMs have been developed for stellar astrophysics \citep[e.g.][]{xiong1978,xiong1986,stellingwerf1982,kuhfuss1986,kuhfuss1987,canuto1992,canuto1993,canuto1997,canuto2011,canuto1998,li2007}. Due to the non-linearity of the Navier-Stokes-equation, higher order terms appear in the equations of the TCM which cannot be computed consistently within the set of equations. To compute the higher order moments, so-called closure relations need to be applied. These closure relations are one of the major sources of uncertainty for any TCM. Ultimately, these closures could be supplied by 3D simulations \citep{chan1989,chan1996,kupka1999a,kupka2007a,kupka2007b,kupka2007c,kupka2007d,kupka2008,viallet2013,arnett2015}.

In this work, we present the results of a TCM applied in a stellar evolution code. We have implemented the TCM by \cite{kuhfuss1987} into the GARching STellar Evolution Code \citep[GARSTEC, ][]{weiss2008}. The additional partial differential equations of the TCM are solved simultaneously with the stellar structure equations with the commonly used Henyey method \citep{flaskamp2003}. Using this implementation, we compute stellar evolution models of low and intermediate mass main-sequence stars from $1.5$ to $8\,M_\odot$ and study their convective cores. For the present paper, surface convection zones are computed with conventional MLT, but models using the Kuhfu{\ss} TCM also for envelope convection are already under way. We demonstrate that the original description of the Kuhfu\ss~convection model leads to erroneous convective properties. In \citet[][hereafter Paper~I]{paper1}, we have shown that the dissipation rate of the kinetic energy plays an important role for the description of turbulent convection. We implemented the description of Paper~I and here we further show that it leads to physically reasonable properties of convection in the framework of the Kuhfu\ss~convection model, which we will briefly introduce and review in Sect.~\ref{sec:theory}. Its application in stellar models is the subject of the following
Sect.~\ref{sec:models}. In Sect.~\ref{sec:varying_masses} we compute stellar models in a mass range from 
1.5~$M_\odot$ to 8~$M_\odot$ with the new model and compare its results with previously used models.
Our discussion in Sect.~\ref{sec:discussion} analyses the origin of the structural differences between
the new 3-equation model and the standard 1-equation model of \cite{kuhfuss1987} and reviews constraints
on core sizes obtained by other available methods. In our conclusions in Sect.~\ref{sec:conclusions} we
provide a summary of the advantages and limitations of the new model and an outlook on further developments.

\section{Implementation of the \cite{kuhfuss1987} model}
\label{sec:theory}
We have implemented the \cite{kuhfuss1987} convection model as described in Appendix~A of Paper~{\sc I} into GARSTEC. This includes the three partial differential equations for the turbulent kinetic energy (TKE) $\omega$, the convective flux $\Pi$, and the entropy fluctuations $\Phi$, as well as the increased dissipation rate in the overshooting zones. We will refer to this model as the 3-equation model. For completeness, we here repeat the final model equations:
\begin{align}
\text{d}_t\omega&=\frac{\nabla_\mathrm{ad}T}{H_p}\Pi-\frac{C_D}{\Lambda}\omega^{3/2}-\mathcal{F}_\omega\label{eqKuh1}\\
\text{d}_t\Pi&=\frac{2\nabla_\mathrm{ad}T}{H_p}\Phi+\frac{2c_p}{3H_p}(\nabla-\nabla_\mathrm{ad})\omega-\mathcal{F}_\Pi-\frac{1}{\tau_\text{rad}}\Pi\label{eqKuh2}\\
\text{d}_t\Phi&=\frac{c_p}{H_p}(\nabla-\nabla_\mathrm{ad})\Pi-\mathcal{F}_\Phi-\frac{2}{\tau_\text{rad}}\Phi\label{eqKuh3}
\end{align}
where $\nabla$ and $\nabla_\mathrm{ad}$ refer to the actual and adiabatic temperature gradient, respectively. The substantial derivative 
is defined as $\mathrm{d}_t=\partial_t+\overline{\vec v}\cdot\nabla$. The radiative dissipation timescale is given as
\begin{align*}
\tau_\text{rad}=\frac{c_p\kappa\rho^2\Lambda^2}{4\sigma T^3\gamma_R^2}.
\end{align*}
Non-local fluxes $\mathcal{F}_a$ are modelled as:
\begin{align*}
\mathcal{F}_a & = \frac{1}{\overline{\rho}}\operatorname{div}\left( - \alpha_a\,\overline{\rho}\,\Lambda\sqrt{\omega}\,\nabla\,\overline{a}\right)
\end{align*}
for $a=\omega,\Pi,\Phi$. The symbols $C_D, \gamma_R, \alpha_\omega, \alpha_\Pi$ and 
$\alpha_\Phi$ are parameters. \cite{kuhfuss1987} suggests a value of $C_D = 8/3\cdot\sqrt{2/3}$ 
and $\gamma_R=2\sqrt{3}$ to be compatible with MLT in the flavour of \cite{boehm1958} in the local 
limit of the model. The length scale of TKE dissipation is denoted by $\Lambda$ 
and will be discussed in detail below. Furthermore, $c_p$ refers to the specific heat capacity at constant 
pressure, $\kappa$ to Rosseland opacity and $\sigma$ to the Stefan-Boltzmann-constant. The variables 
$T$, $\rho$, and $H_p$ are temperature, density, and pressure scale height, as usual in stellar structure 
models. For the details of the derivation and the definitions of the symbols, we refer to the appendix A in 
Paper~{\sc I} and to \cite{kuhfuss1987} and \cite{flaskamp2003}.

\subsection{The 1-equation model}
\label{sec1Eq}
In addition to the 3-equation model, \cite{kuhfuss1987} has also suggested a simplified version of his TCM \citep[see also][]{kuhfuss1986}. The number of equations is reduced to one by introducing the following approximation for the convective flux:
\begin{align}
	\Pi=-\alpha_s\Lambda\sqrt{\omega}\frac{c_p}{H_p}(\nabla-\nabla_\mathrm{ad})\,.
	\label{eqconvflux1}
\end{align}
The parameter $\alpha_s=(1/2)\sqrt{2/3}$ has been calibrated to MLT and $\Lambda$ is again 
the length-scale of TKE dissipation. As for the dissipation parameter $C_D$, 
the parameter value of $\alpha_s$ is obtained by calibrating convective velocities and fluxes 
of the local 1-equation model to MLT analytically. This approximation of the convective flux allows eliminating 
the convective flux from the $\omega$-equation such that it is only necessary to solve a single equation. 
We will refer to this model as the 1-equation model. When applying the 1-equation model, the length scale 
for the dissipation of the TKE will be defined as $\Lambda=\alpha H_p$ -- with a freely adjustable 
parameter $\alpha$ -- throughout the rest of the paper, equivalent to the usual mixing length. 
For a value of order unity for $\alpha$ results are indeed comparable to MLT results.
%
\subsection{Convection equations}
As most modern stellar evolution codes GARSTEC makes use of the implicit Henyey scheme to solve the four stellar structure equations \citep{henyey1964,henyey1965,kippenhahn1967}. The equations describing convection by MLT are solved algebraically outside the four stellar structure equations. To incorporate the three equations describing the convection model, \cite{flaskamp2003} has extended the Henyey-scheme of GARSTEC to solve for in total seven variables (four stellar structure variables + three convection variables).
A solution for both the stellar structure and the convective variables is found by iterating over all variables simultaneously. The coefficients of the convection model depend on the stellar structure variables, such that the behaviour of the convection model is strongly coupled to the stellar structure. On the other hand, the stellar structure is coupled to the convective variables through the temperature gradient and the chemical composition. As described in Paper~{\sc I} the temperature gradient of the stellar model is computed self-consistently from the convective flux in each iteration (see Eq.~A.7 in Paper~{\sc I}). The chemical mixing in convective zones is computed in the framework of a diffusion equation, alongside the composition changes due to nuclear burning after the structure equations have been solved. The diffusion constant is computed from the TKE determined by the convection model. Following \cite{langer1985} the diffusion coefficient is computed as:
\begin{align}
    D=\alpha_s\Lambda\sqrt{\omega}
    \label{eqdiffcoeff}
\end{align}
where we have chosen the same parameter $\alpha_s$ as for the diffusion coefficient of entropy in Eq.~(\ref{eqconvflux1}).

The equations can also describe time-dependent effects. This was demonstrated for example by \cite{flaskamp2003}, when computing models through the core helium flash at the tip of the red-giant branch, by \cite{wuchterl1998} in an application to protostars and non-linear pulsations, and by \cite{feuchtinger1999b} for RR Lyrae stars. In this work, however, we focus on main-sequence stars which evolve on the nuclear timescale of hydrogen burning. This means that structural changes are sufficiently slow to neglect time-dependent terms and immediately solve for the stationary solution of the convection equations (left-hand sides of the TCM equations Eq.~(\ref{eqKuh1})-(\ref{eqKuh3}) are set to zero). By iterating for the stationary solution, the code searches for the converged stellar structure and convection variables for a given chemical composition. When non-local effects are included in the convection model, the stationary solution describes the  overshooting zone self-consistently. Its extent and temperature gradient are only constrained by the convection model, without any external descriptions.

As described in Paper~{\sc I} and at the beginning of this section, the \cite{kuhfuss1987} convection model contains a number of parameters. The values for these parameters need to be set. As described above,  $C_D$ and $\gamma_R$ are obtained by calibrating a local model to MLT. The parameter values for the non-local terms $\alpha_\omega,\alpha_\Pi$ and $\alpha_\Phi$ cannot be calibrated to MLT as they describe intrinsically non-local effects. \cite{kuhfuss1987} suggests a default value of $\alpha_\omega=0.25$ by comparing kinetic energy and dissipation in a ballistic picture. No default values have been given for the parameters $\alpha_\Pi$ and $\alpha_\Phi$ in the non-local case. Although both the 1- and 3-equation models still contain a number of parameters, they are advantageous compared to for example MLT because the parameters describe physically more fundamental properties of the theory. For example, the parameter $\alpha_\omega$ describes the impact of the non-local flux of the TKE, which is responsible for the extent of the overshooting region. However, compared to diffusive overshooting or step overshooting $\alpha_\omega$ does not set the actual length-scale of the overshooting. The extent of the overshooting is determined self-consistently from the solution of the TCM equations. We will investigate the impact of these other parameters further below.
%
\subsection{Dissipation rate}
In Paper~{\sc I}, we showed that the original description of the dissipation rate proposed by \cite{kuhfuss1987} 
leads to an excessively large overshooting region. Therefore, in Paper~{\sc I} the dissipation rate was increased 
by taking into account buoyancy waves as a sink for the TKE. The increase of the dissipation rate 
is realised through a modification of 
its associated dissipation length scale. The dissipation rate is inversely proportional to this length scale, 
$\epsilon = c_{\epsilon} \omega^{3/2} / \Lambda$, such that a decrease of the latter leads to an increase 
of the dissipation rate. This modification of the TKE dissipation length scale was implemented through a harmonic sum:
\begin{align}
\frac{1}{\Lambda}=\frac{1}{\alpha H_p}+\frac{1}{\beta_s r}
\label{eqdissold}
\end{align}
where the newly introduced parameter $\beta_s$ is defined as:
\begin{align*}
\beta_s=(1+\lambda_s\tilde N)^{-1}
\end{align*}
and
\begin{align}
\lambda_s=c_4\Lambda \omega^{-1/2}
\label{eqlambdas}
\end{align}
where $c_4=c_3/(c_2c_\epsilon)$ (see Eq.~21 in Paper~{\sc I}). The parameters $c_2=1.92$ and $c_3=0.3$ are model parameters 
from \cite{canuto1998} and $c_\epsilon$ is the dissipation parameter of the convection model. 
The buoyancy frequency $\tilde{N}$ is computed according to 
\begin{align*}
\tilde{N}^2=\frac{g^2\rho}{p}\left(\nabla_\text{ad}-\nabla+\nabla_\mu\right)
\end{align*}
assuming an ideal gas law. Here, $\nabla_\mu$ indicates the dimensionless mean molecular weight gradient 
and $g$ refers to the gravitational acceleration.  Close to the stellar centre the pressure scale height 
and in turn the dissipation length scale $\Lambda$, if defined through the pressure scale height, diverge. 
To avoid this divergence, \cite{flaskamp2003} introduced the modification by \cite{wuchterl1995} in the Kuhfu\ss~model. 
This extension of the model we also implemented equivalently to Eq.~(\ref{eqdissold}), but with a constant parameter $\beta_s=1$. 
In unstably stratified regions, the new correction factor is not applied and $c_4=0$ as $c_3$ drops to 0 \citep{canuto1998}. 
Hence, the harmonic sum Eq.~(\ref{eqdissold}) recovers the \cite{wuchterl1995} expression automatically. 
We note that at this point where $\nabla-\nabla_\text{ad}\to0$, also $\tilde{N}\to0$ and $\beta_s$ smoothly transitions to 1, 
such that setting $c_3=0$ does not introduce any discontinuity. The harmonic sum Eq.~(\ref{eqdissold}) 
can be converted into an equation for the dissipation length $\Lambda$. Rewriting Eq~(\ref{eqdissold}) yields:
\begin{align}
\Lambda=\underbrace{\frac{r}{r+\alpha H_p\frac{1}{\beta_s}}}_{<1}\alpha H_p\label{eqlambdared}
\end{align}
which immediately shows that the derived expression is in essence a reduction factor for the mixing length. 
Plugging in the definitions for $\beta_s$ and $\lambda_s$  one finds the following expression:
\begin{align*}
\Lambda=\frac{r}{r+\alpha H_p\left(1+c_4\Lambda \omega^{-1/2}\tilde N\right)}\alpha H_p
\end{align*}
This is a quadratic equation in $\Lambda$ which can be solved to obtain the reduced length-scale. 
Here, we have expressed the dissipation rate timescale $\tau = 2 \omega / \epsilon$ in terms 
of $\omega$ and $\Lambda$ through noting that $\epsilon = c_{\epsilon} \omega^{3/2} / \Lambda$
and $\tau = 2 \Lambda / (c_{\epsilon} \omega^{1/2})$ which allowed rewriting an expression
proportional to the  ratio of turbulent kinetic energy to buoyancy time scales, $\tau/\tau_{\rm b}$, 
into one proportional to $\Lambda \omega^{-1/2}\tilde N$ (see Paper~{\sc I} for details).
The final model of $\Lambda$ in the convection zone reads:
\begin{align}
\Lambda(r)=-\frac{r+\alpha H_p}{2c_4\tilde N\omega^{-1/2}\alpha H_p}+\sqrt{\left[\frac{r+\alpha H_p}{2c_4\tilde N\omega^{-1/2}\alpha H_p}\right]^2+\frac{r}{c_4\tilde N\omega^{-1/2}}}
\end{align}
for $\nabla<\nabla_\text{ad}$, and 
\begin{align}
\Lambda(r)=\left(\frac{1}{\alpha H_p}+\frac{1}{\beta_{\rm c} r}\right)^{-1}, & \label{eqKFLambda}
\end{align}
for $\nabla>\nabla_\text{ad}$, where $\beta_{\rm c}=1$. To obtain a physically reasonable, positive dissipation 
length scale, $\Lambda$ the plus sign in front of the square root in the solution of the quadratic equation has to be chosen.

We note that the parameter $c_\epsilon$ takes different values in the Canuto and Kuhfu\ss~convection 
models. Here, we take $c_\epsilon=C_D$ which is the dissipation parameter in the 
Kuhfu\ss~model.
In unstably stratified regions with $c_4=0$ Eq.~(\ref{eqdissold}) solves explicitly for $\Lambda$. 
In stably stratified regions, the parameter takes a value of $c_4\approx0.072$ using the parameters 
$c_2$ and $c_3$ from \cite{canuto1998} and $c_\epsilon=C_D$.

As mentioned already in Sect.~3.6 in Paper~{\sc I}, the effect of changing $c_\epsilon$ on changing $c_4$ to some extent  cancels out. We discuss this further in Appendix~\ref{secparamdep}.

%
%
\section{Stellar models}
\label{sec:models}

We have used GARSTEC to compute stellar models in a mass range of 1.5-8~$M_\odot$ in the core hydrogen burning phase. We have used the OPAL equation of state, OPAL opacities \citep{iglesias1996}, extended by low temperature opacities by J.~Ferguson \citep[private communication and][]{ferguson2005}, both for the \citet[][GN93]{grevesse1993} mixture of heavy elements. For the initial mass fractions we have chosen $X=0.7$ and $Z=0.02$ for all models. Convective chemical mixing is described in a diffusive way. We use MLT plus diffusive overshooting as described by \cite{freytag1996} to evolve the models through an initial equilibration to the beginning of the main-sequence phase. Then we first switch to the 1- and subsequently to the 3-equation model to generate a starting model for the computation with the 3-equation model. As we are interested in the effects of overshoot mixing, all models shown in the following are computed, including the non-local terms in the 1- and 3-equation version of the Kuhfu{\ss} theory. 

For the diffusive overshooting, we have used the default GARSTEC parameter value of $f_\text{OV}=0.02$, which has been calibrated by fitting GARSTEC-isochrones to the colour-magnitude diagrams of open clusters \citep{magic2010}. The diffusion coefficient is computed according to Eq.~(\ref{eqdiffcoeffadhoc}). For small convective cores, excessively large overshooting zones can occur when applying the ad~hoc overshooting schemes due to the diverging pressure scale height in the centre. To avoid such unfavourable conditions a reduced overshooting parameter value is determined in GARSTEC, by applying a geometrical cut-off depending on the comparison between the radial extent of the convective region and the scale-height at its border. For a brief discussion of different geometric cut-off descriptions, we refer to Appendix~\ref{seccut}. For the results in this section, the ``tanh''-cut-off according to Eq.~(\ref{eqcuttanh}) has been used. For the parameter $\alpha_\omega$ we have chosen a value of 0.3 as this value results in a similar convective core size as the ad~hoc overshooting models for $f_{\rm OV}=0.02$ in the 5~$M_\odot$ model, which is in the middle of our mass range. For the parameters $\alpha_\Pi$ and $\alpha_\Phi$ we have chosen the same value assuming that the non-local transport behaves similar for all convective variables. We would like to point out that $\alpha_\omega$ is an adjustable parameter and an external calibration will be necessary. This will be discussed further below.

\subsection{The 3-equation model}  \label{sec3eq}
\begin{figure}
	\centering
	\includegraphics{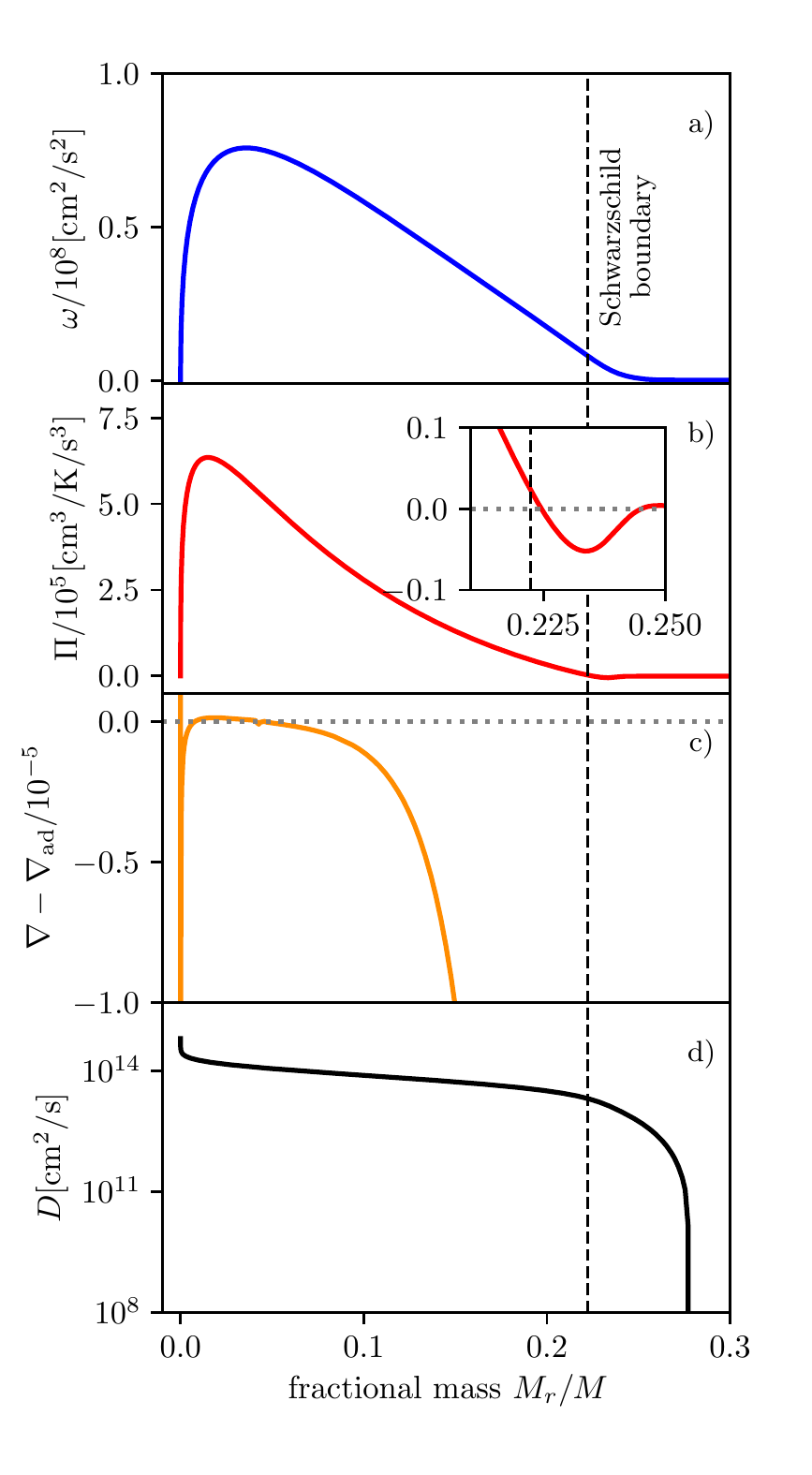}
	\caption{Summary of the interior structure of the convective core of a $5\,M_\odot$ main-sequence model, calculated with the 3-equation model. The black dashed line indicates the Schwarzschild boundary. The different panels show a) TKE b) convective flux variable c) super adiabatic temperature gradient and d) the diffusion coefficient according to Eq.~(\ref{eqdiffcoeff}). The selected stellar model has a central hydrogen abundance of $X_{\rm c}=0.6$. The inset in panel b) shows the region of negative convective flux just beyond the Schwarzschild boundary.}
	\label{figsummary}
\end{figure}
As a representative example, we will first discuss the evolution and the internal structure of a 5~$M_\odot$ model applying the 3-equation, non-local convection theory. Figure~\ref{figsummary} shows the profiles of the TKE variable $\omega$ (panel a), the convective flux variable $\Pi$ (panel b), the superadiabatic gradient $\nabla-\nabla_\text{ad}$ (panel c), where $\nabla$ is the temperature gradient resulting from the convection model, and the diffusion coefficient according to Eq.~(\ref{eqdiffcoeff}) on a logarithmic scale (panel d). The TKE clearly extends beyond the Schwarzschild boundary, which we compute as usual as the point where $\nabla_\text{ad}=\nabla_\text{rad}$. 
Such a behaviour cannot be observed in MLT models, as it is a direct result of the non-local terms in the Kuhfu\ss~convection model. The associated diffusion coefficient shows a rather high value beyond the Schwarzchild boundary and throughout the overshooting zone. 
This will increase the size of the 
mixed convective core and therefore naturally create an overshooting zone. Given the profile of the diffusion coefficient, the chemical mixing will resemble the step overshooting rather than the diffusive overshooting scheme. The convective flux variable shows a region of negative flux 
beyond the Schwarzschild boundary. In the lower panel of Fig.~\ref{figcompareMLT} the region of negative convective flux 
is shown enlarged in the inset (likewise for the convective flux variable $\Pi$ in the middle panel of Fig.~\ref{figsummary}).
This is due to the braking of the convective motions in the stable, radiative stratification. The extent and magnitude of the 
negative convective flux are very comparable to early results from \cite{xiong1986}, who found that the convective flux penetrates 
less deeply into the stable layers than for example the kinetic energy and has a nearly negligible magnitude compared to the total flux.

Finally, panel c) shows the superadiabatic temperature gradient of this stellar model (notice the vertical scale). 
In the inner part of the convection zone the model shows a very small superadiabatic gradient as it is expected 
for regions with convective driving. At a fractional mass of about 0.05 the temperature gradient drops below 
the adiabatic value. In contrast to local models this point does not coincide with the formal Schwarzschild boundary, 
however, the sign change happens substantially before the formal boundary. At the formal Schwarzschild boundary, 
the temperature gradient has dropped to about $\mathcal{O}(10^{-3})$ below the adiabatic value. The comparison of panels b) and c) 
shows that there exists an extended region in the model in which the convective flux is positive while the temperature gradient 
is already subadiabatic. This region is also known as a Deardorff layer \citep{deardorff1966} and has been observed in simulations of 
stellar convection \citep{chan1992,muthsam1995,muthsam1999,tremblay2015,kapyla2017,kupka2018} or other Reynolds stress models 
\citep{kupka1999b,xiong2001,kupka2002,montgomery2004,zhang2012a}. 

Such a layer cannot exist in convection models, 
which do not have enough degrees of freedom. In MLT and the Kuhfu\ss~1-equation model, the convective flux is directly coupled to 
the superadiabatic gradient and the convective velocities (see Eq.~\ref{eqconvflux1} for the 1-equation model). This of course inhibits 
any region in which the convective flux and the superadiabatic gradient have a different sign like in the Deardorff layer and forces 
the convective flux and the TKE to have the same penetration depth. The 3-equation model lifts 
the strong coupling of convective flux and superadiabatic gradient by directly solving for two more variables ($\Pi$ and $\Phi$) 
and therefore allows for the existence of such a layer. \cite{deardorff1966}, in the case of atmospheric conditions, argued
that it is mainly the non-local term in the equation for $\Phi$ which supports the positive heat flux for subadiabatic temperature 
gradients \citep[see also Sec. 13 of][and references therein]{canuto1992}.  The diffusion term $\mathcal{F}_\Phi$ in the equation for the entropy fluctuations Eq.~(\ref{eqKuh3}) acts as a source of entropy fluctuations even though no local source (a superadiabatic temperature gradient) is present. This allows for the outward directed transport of entropy fluctuations, that is a positive convective flux even in subadiabatic layers. This has little impact on the stellar structure and evolution, as the temperature gradient remains nearly adiabatic, and the whole convection zone is chemically well mixed. The existence of such a layer is therefore expected, and confirms the physical relevance of the 
3-equation Kuhfu\ss~model. The extent of this layer is difficult to determine from a priori arguments, and asteroseismic analyses might provide observational constraints in the future.
\begin{figure}
	\centering
	\includegraphics{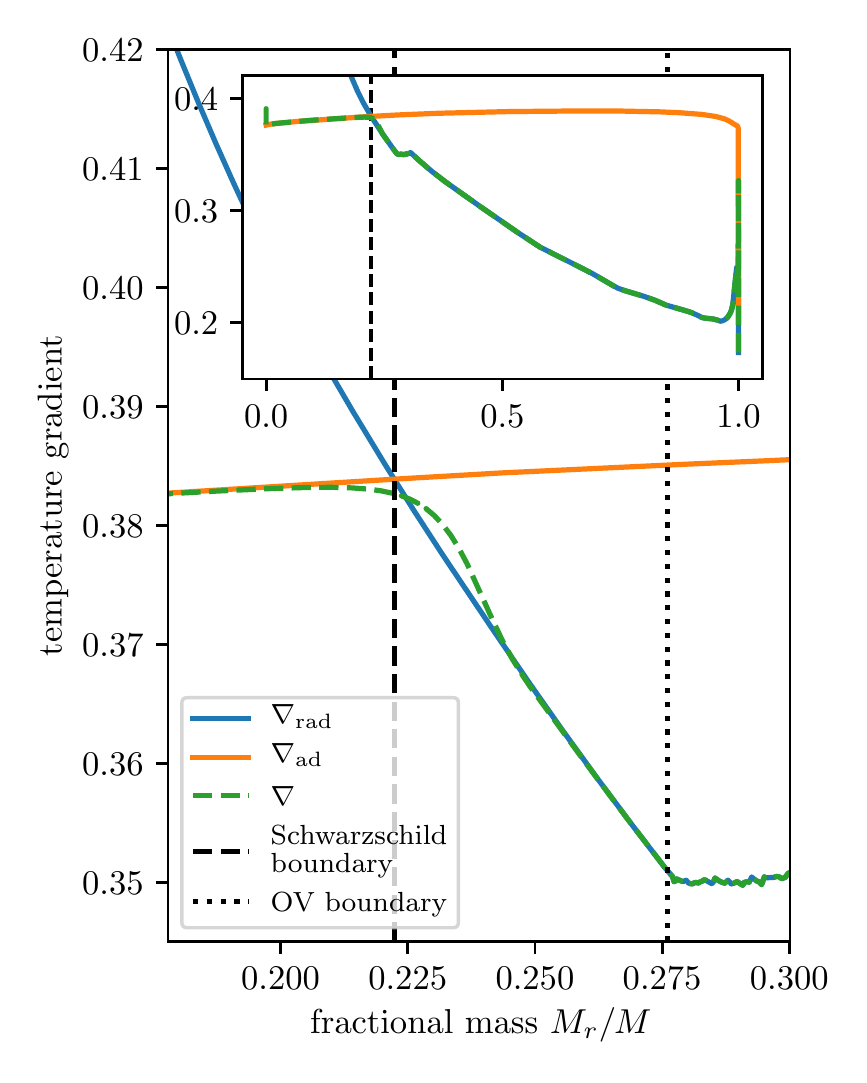}
	\caption{Temperature gradients in the overshooting zone of the same $5\,M_\odot$ main-sequence model as in Fig.~\ref{figsummary}. The blue and orange lines indicate the radiative ($\nabla_\text{rad}$) and adiabatic ($\nabla_\text{ad}$) temperature gradients, respectively. The green dashed line indicates the model temperature gradient $\nabla$ obtained from the 3-equation non-local convection model. The black dashed line indicates the Schwarzschild boundary, while the black dotted line indicates the boundary of the well mixed overshooting region. The inset shows the three temperature gradients from the centre to the surface of the stellar model. The selected stellar model has a central hydrogen abundance of $X_{\rm c}=0.6$.}
	\label{fignabla}
\end{figure}

In Fig.~\ref{fignabla} we show the temperature gradients in the overshooting zone of the same $5\,M_\odot$ main-sequence model as in Fig.~\ref{figsummary}. At the formal Schwarzschild boundary the model temperature gradient has already a slightly subadiabatic value as discussed previously. Beyond the Schwarzschild boundary, the model temperature gradient does not drop to the radiative gradient immediately. Instead, it gradually transitions from slightly subadiabatic to radiative values in a rather narrow mass range. As a consequence, the model temperature gradient takes slightly super-radiative values in this transition region. However, the temperature gradient reaches a radiative value well before the boundary of the mixed region, indicated with the black dotted line in Fig.~\ref{fignabla}. Considering the small extent of the super-radiative region and the small deviation from the radiative temperature gradient, the overshooting zone in the 3-equation non-local model is mostly radiative. We would like to point out that the shape of the temperature gradient is not subject to assumptions about the thermal stratification (e.g. adiabatic, radiative or any gradual transition between both) in the overshooting zone but instead is a result of the convection model. In the transition region, the convective flux is negative due to the buoyancy braking, which effectively means that energy transport by convection is directed inwards instead of outwards. This effect is counter-balanced by increasing the energy transport by radiation, through an increased model temperature gradient \citep[e.g.\ ][]{chan1996}.

The temperature gradient of the 3-equation model is comparable to results of different TCM approaches \citep{xiong2001,li2007} for the base of the solar convective envelope. Both \cite{zhang2012a} (their Figs.~6 and 7) and \cite{xiong2001} (their Fig. 8) find a temperature gradient that transitions gradually from the adiabatic to the radiative value. They also find a Deardorff layer with a degree of subadiabaticity at the formal Schwarzschild boundary comparable to our findings. From the convective flux as presented in \cite{xiong1986} one also would expect a similar temperature gradient in the overshooting zone. Furthermore, the shape of the model temperature gradient is also in qualitative agreement with the discussion in \cite{viallet2015}. They argue that under the physical conditions in convective cores, in regions of overshooting efficient chemical mixing and a gradually transitioning temperature gradient are expected. In the 3-equation non-local model, the extent of the nearly adiabatic overshooting zone is controlled by the shape of the negative convective flux in the overshooting zone. For smaller (more negative) values of the convective flux (i.e.\ more efficient buoyancy braking) the temperature gradient is expected to be closer to the adiabatic value, while for larger (less negative) values it will be closer to the radiative temperature gradient. In Eq.~(\ref{eqKuh1}) the negative convective flux and the dissipation term act as sink terms  in the overshooting zone. Hence, the behaviour of the dissipation term will impact also on the convective flux and in turn on the value of the temperature gradient in the overshooting zone. In computations with the 1-equation non-local version of the theory, the negative convective flux is the dominant sink term for the TKE and the actual dissipation term is negligible (see Fig.~8 of Paper~{\sc I}). This leads to more negative values of the convective flux and thus to a mostly adiabatic temperature gradient in the overshooting zone. 
We will discuss this in more detail in Sec.~\ref{sepeclet} (see also Fig.~\ref{figpecgrad}).

\begin{figure}
	\centering
	\includegraphics{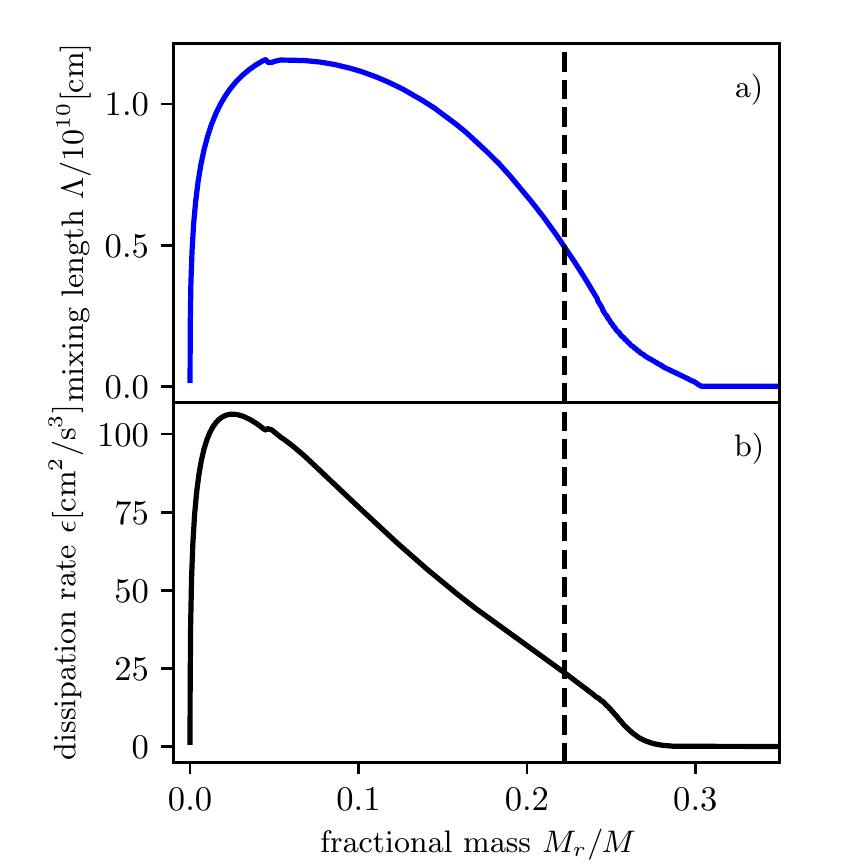}
	\caption{Dissipation length scale $\Lambda$ (a) and dissipation rate (b) in a $5\,M_\odot$ main-sequence model. Same stellar model as in Fig.~\ref{figsummary}. The vertical dashed line indicates the Schwarzschild boundary of the model.}
	\label{figdissipation}
\end{figure}

As discussed above and in Paper~{\sc I} we have implemented the increase of the dissipation rate in the overshooting zone 
through a decrease of the dissipation length scale $\Lambda$. In the original version, \cite{kuhfuss1987} models the dissipation 
of the TKE by a Kolmogorov-type term
($\epsilon=C_D  \omega^{3/2} / \Lambda$ with $\Lambda(r)$ as in Eq.~\ref{eqKFLambda}).
The dissipation length scale $\Lambda$ describes the scale over which the kinetic energy is dissipated such that in 
the Kolmogorov model, at fixed TKE, a shorter length scale results in an increased dissipation rate. In Fig.~\ref{figdissipation} 
we show the length scale $\Lambda$ (panel a) and the dissipation rate (panel b) computed in the same stellar model as in 
Fig.~\ref{figsummary}. At a fractional mass of about 0.05 the profile of the dissipation length scale shows a slight kink in this 
model, where the transition starts. At about 0.28 in fractional mass, the dissipation length scale $\Lambda$ 
drops to zero, which means that convective motions stop at that point. This coincides with the dying of the TKE beyond 
the Schwarzschild boundary, as seen in Fig.~\ref{figsummary}, panel a). The onset of the decrease of the dissipation 
length scale $\Lambda$ also coincides with the sign change of the superadiabatic gradient (see Fig.~\ref{figsummary}, panel c). Towards the 
centre of the model, the dissipation length scale $\Lambda$ drops to zero as well. This is a result of the \cite{wuchterl1995} 
correction, also used in our implementation (cf.~Paper~{\sc I}). We have also investigated the
prescription of \cite{roxburgh2007} as an alternative to that one of \cite{wuchterl1995} 
but found little difference with respect to the overshooting region (see also Paper~{\sc I}). 
\paragraph{Parameter sensitivity.}
We explore some of the parameter dependencies in Appendix~\ref{secparamdep}. We studied the impact of the new parameter $c_4$, appearing in Eq.~(\ref{eqlambdas}) (see also Eq.~21 in Paper~{\sc I}), on the structure of the convective core. A comparison of TKE profiles for different values of $c_4$ is shown in Fig.~\ref{figvariation}. The comparison demonstrates that although the parameter has some impact on the overshooting extent, the main properties of the model are not changed substantially. The parameter $\alpha_\omega$ , appearing in Eq.~(\ref{eqKuh1}), controls the non-local flux of the turbulent kinetic energy. Because this flux is mainly responsible for the extension of the kinetic energy beyond the Schwarzschild boundary, one expects that this parameter impacts on the overshooting distance. Figure~\ref{figovcomparison} shows a comparison of different hydrogen profiles of a 5~$M_\odot$ star computed with different values of $\alpha_\omega$. For a higher value of the parameter, the size of the convective core is larger throughout the evolution of the model. Smaller parameter values reduce the convective core size. Although \cite{kuhfuss1987} provides a default value for $\alpha_\omega$, this value is not known a priori from the theory. Hence, a calibration will be necessary, for example from observations or 3D hydrodynamic simulations. The parameters of the non-local terms in the $\Pi$ and $\Phi$ equations $\alpha_\Pi$ and $\alpha_\Phi$, appearing in Eq.~(\ref{eqKuh2}) and~(\ref{eqKuh3}) respectively, have a negligible impact on the overshooting extent (see Fig.~\ref{figovcomparisonPiPhi} in the appendix). Instead, they have a larger impact on the temperature gradient as the variables $\Pi$ and $\Phi$ are more closely related to the temperature structure. By increasing the parameter $\alpha_\Pi$, the magnitude of the superadiabatic gradient increases, while the extent of this region stays the same. Decreasing the value of the parameter $\alpha_\Phi$ is increasing the size of the superadiabatic region, therefore reducing the size of the Deardorff layer, while the magnitude stays the same. This is expected because the non-local term in the $\Phi$ equation is the one which is driving convection in the Deardorff-layer.  

\subsection{Comparison to MLT}
\begin{figure}
	\centering
	\includegraphics{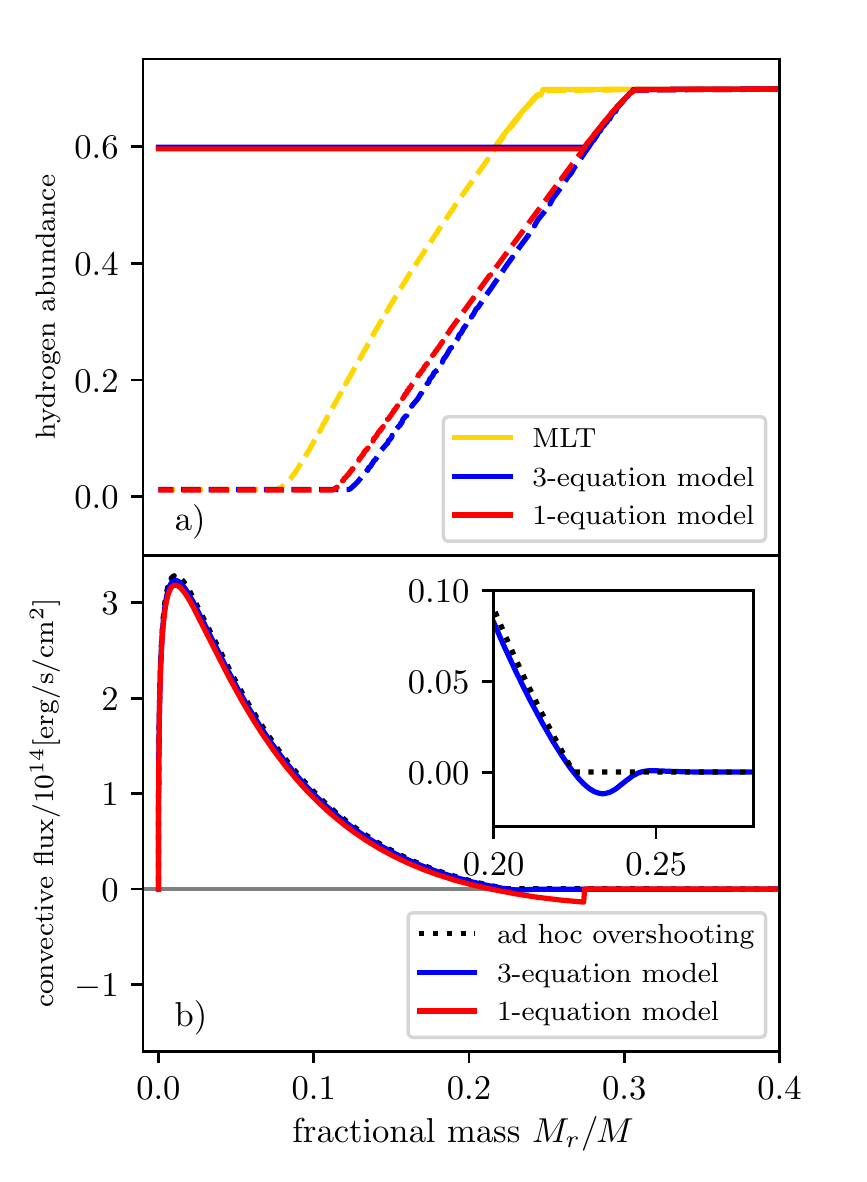}
	\caption{Comparison of models (yellow lines) using MLT without overshooting to those computed with the 3-equation theory (blue lines) and the 1-equation model (red lines). Panel a) compares the hydrogen profiles at an early stage on the main sequence, when $X_{\rm c}=0.6$ (solid lines), and at the end of it ($X_{\rm c}\approx0$; dashed lines). Panel b) shows the convective fluxes of an MLT model with diffusive overshooting (dotted black line), a 3-equation model (solid blue line) and a 1-equation model (solid red line). These models have been selected to have the same central hydrogen abundance of $X_{\rm c}=0.6$ and the same chemically homogeneous core size.}
	\label{figcompareMLT}
\end{figure}

Mixing-length theory is still the most commonly used theory to describe convection in stars. Therefore, we will now compare the results of the 3-equation model with results obtained from standard MLT, that is\ without and with an additional treatment of overshooting. The fractional hydrogen abundances early on and at the end of the main sequence computed with the 3-equation model and an MLT model without overshooting are shown in the upper panel of Fig.~\ref{figcompareMLT}. At the same central hydrogen abundance, the fully mixed region of the 3-equation model always extends past that of the MLT model. This is comparable to models which include ad~hoc overshooting beyond the formal Schwarzschild boundary. In contrast to the ad~hoc overshooting models, the overshooting in the 3-equation model results from the solution of the model equations. The lower panel of Fig.~\ref{figcompareMLT} shows a comparison of the convective fluxes in the 3-equation model and an MLT model. Both models have been selected to have the same central hydrogen abundance. Ad~hoc overshooting has been included in this MLT model, and tuned in such a way as to obtain a model with the same core size as obtained from the 3-equation model. In the bulk of the convection zone, both fluxes show very close agreement. Beyond the Schwarzschild boundary, the narrow region with negative convective flux in the non-local model can be identified. This region is shown enlarged in the inset.

\begin{figure}
	\centering
	\includegraphics{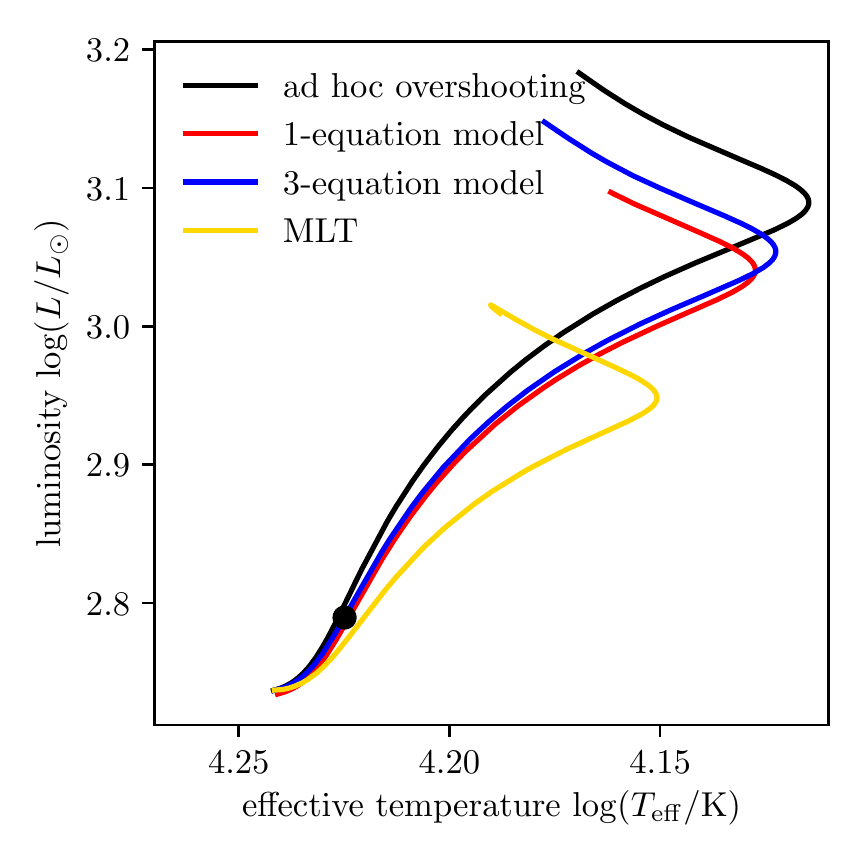}
	\caption{Evolutionary tracks in the Hertzsprung-Russell diagram of a $5\,M_\odot$ model computed with MLT, shown with a yellow line, ad~hoc overshooting, shown with a black line, and the 3-equation, non-local model, shown with a solid blue line. The black dot marks the model selected at a central hydrogen abundance of $X_{\rm c}=0.6$.}
	\label{figHRDcompare}
\end{figure}

Finally, in Fig.~\ref{figHRDcompare} we compare the evolutionary tracks of the 3-equation model, indicated by the solid blue line, with an MLT model without ad~hoc overshooting (yellow line) and an MLT model including ad~hoc overshooting (black line). The black dot indicates the position of the stellar model with $X_{\rm c}=0.6$ discussed in subsection~\ref{sec3eq} in Figs.~\ref{figsummary} to \ref{figcompareMLT}. The computation of the 3-equation model starts at the beginning of the main sequence from an MLT model including diffusive overshooting as described above and then evolves through core hydrogen burning up until core hydrogen exhaustion. Compared to the MLT model, the non-local model shows a higher luminosity throughout the main sequence, as expected for the larger convectively mixed core.\\

\subsection{Comparison to the 1-equation non-local models}
\begin{figure}
\centering
 \includegraphics[]{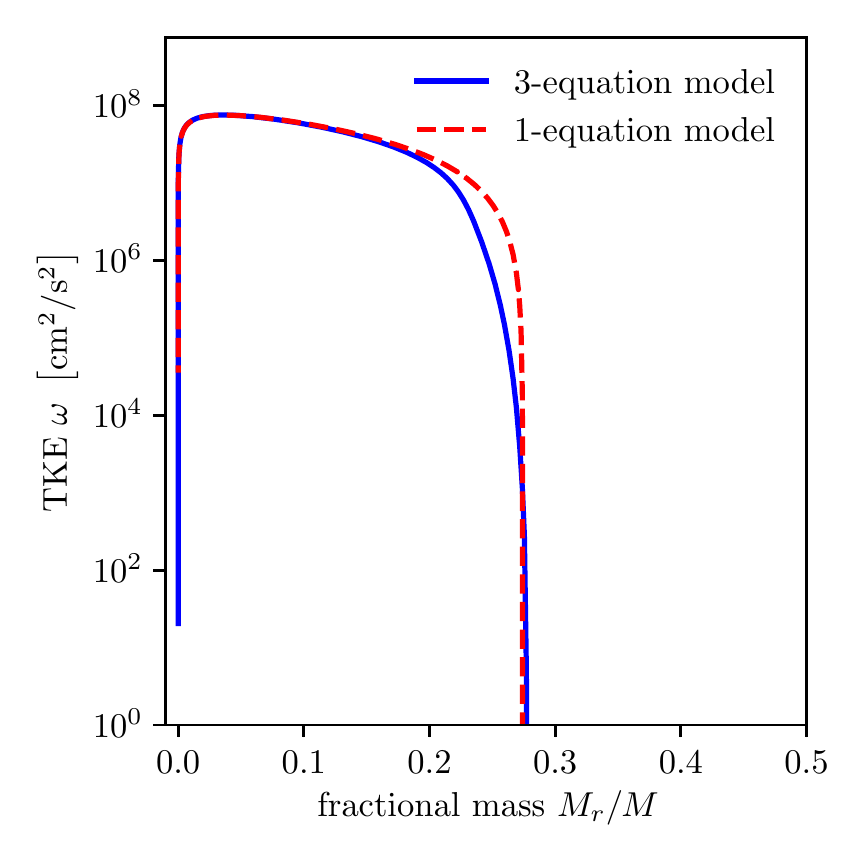}
 \caption{Comparison of the TKE in the 1-equation non-local model and the 3-equation non-local model with enhanced dissipation rate of TKE in a 5~$M_\odot$ main-sequence model on a logarithmic scale. The models have been selected to have the same central hydrogen abundance as the model in Fig.~\ref{figsummary}. The 1-equation model computes $\Lambda$ according to Eq.~(\ref{eqKFLambda}).}
 \label{fig1Eqcompare}
\end{figure}

In addition to the 3-equation non-local models, we have also computed stellar models, in which convection is described by the 1-equation model (see Sec.~\ref{sec1Eq}). 
As before, we included the non-local terms. Figure~\ref{fig1Eqcompare} shows a comparison of the TKE profiles for the 1- and 3-equation, non-local models 
on a logarithmic scale. The TKE on a linear scale can be found in Fig.~\ref{fig1Eqcompareuntrans} in the appendix. The models have been selected at 
the same central hydrogen abundance to ensure that they are in the same evolutionary stage. This shows that the overshooting extent in the 1-equation non-local model
is very comparable to the 3-equation non-local model for the same choice of the parameter $\alpha_\omega$. The overall behaviour of the 1-equation non-local model and 
the new 3-equation non-local model looks very similar. The overshooting extent is clearly limited, and there is a steep drop in the TKE at the overshooting 
boundary. Also, the absolute values of the TKE look comparable in the bulk of the convection zone (Fig.~\ref{fig1Eqcompareuntrans}). We will analyse 
the absolute value of the TKE in more detail below. We note that this result is obtained without tuning the parameters of the models. For the parameters 
which both models have in common, the same values were chosen. In the overshooting zone, the 3-equation model has much smaller TKE than the 1-equation 
model. Nevertheless, the energies are still high enough to fully mix the overshooting region in the 3-equation model.

In the upper panel of Fig.~\ref{figcompareMLT} we compare the hydrogen profiles of the 1-equation model with the results from the 3-equation model and an MLT model at the beginning and at the end of the main-sequence. As expected from the similar TKE profiles shown in Fig.~\ref{fig1Eqcompare} the hydrogen profile of the 1-equation model extends past the local MLT model and looks very similar to the 3-equation model. Towards the end of the main-sequence, the 1-equation model has a smaller core than the 3-equation model, leading to a slightly different slope in the hydrogen profiles. Likewise, the evolutionary track shown in Fig.~\ref{figHRDcompare} of the 1-equation model looks very similar to the 3-equation model. Towards the end of the main-sequence, the luminosity is slightly lower owing to the smaller convective core. In the lower panel of Fig.~\ref{figcompareMLT} we compare the convective flux of the 1-equation model to the 3-equation model and an MLT model including ad~hoc overshooting. As for the 3-equation model, the convective flux shows close agreement with the other two models in the bulk of the convection zone. Only in the overshooting zone, where the convective flux becomes negative, discrepancies become apparent. Compared to the 3-equation model, the zone of negative convective flux is more extended and the absolute value is larger, that is the convective flux is more negative. This can be attributed to the parametrisation of the convective flux in the 1-equation model according to Eq.~(\ref{eqconvflux1}). As discussed in Sec.~4 of Paper~{\sc I} (see also their Fig.~8) the buoyancy term, proportional to the convective flux, acts as the main sink term in the overshooting zone, requiring larger absolute values in the overshooting zone. We would like to note that this strongly negative convective flux in the overshooting zone will also cause the 1-equation model to have a nearly adiabatic overshooting zone, as compared to the nearly radiative overshooting zone in the 3-equation model.

\section{Non-local convection for varying initial masses} \label{sec:varying_masses}
\begin{figure}
\centering
 \includegraphics[]{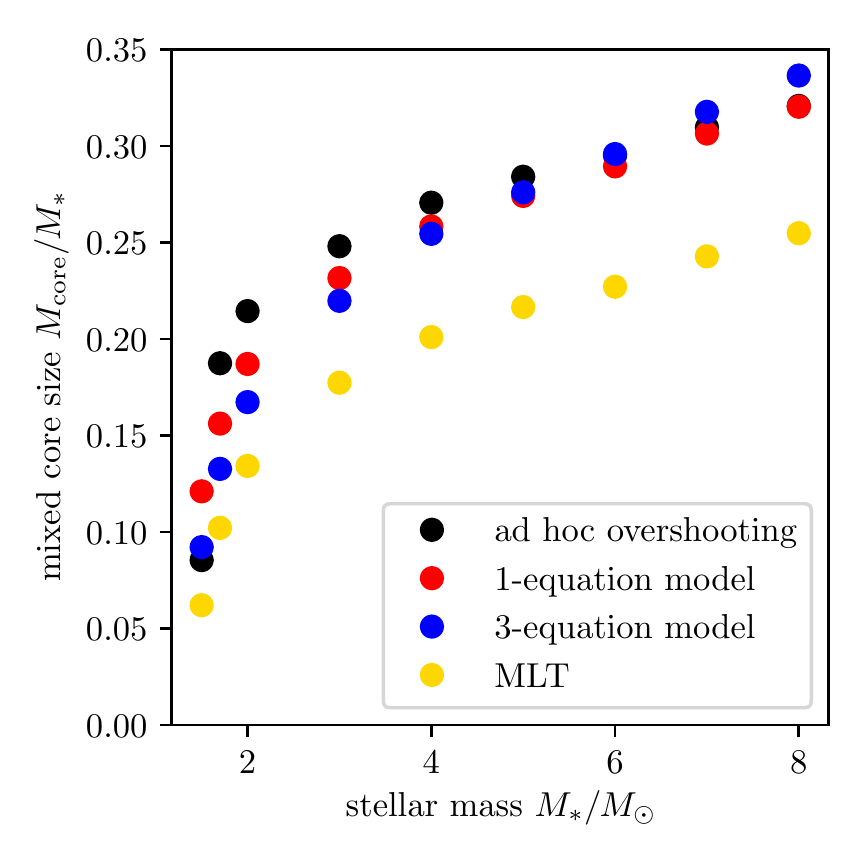}
 \caption{Comparison of mixed core sizes of stellar models in units of the stellar mass $M_*$ over a range of initial stellar masses computed with different convection models. The models with ad~hoc overshoot include a geometric cut-off to limit the size of small convective cores in lower mass stars. The models have been selected at a central hydrogen abundance of $X_{\rm c}=0.6$. The MLT models are computed without overshooting.}
 \label{figmixedcorecompare}
\end{figure}

We computed stellar models in a mass range of 1.5-8~$M_\odot$, using the 3-equation non-local model. The models have been constructed in the same way and using the same parameters as for the 5~$M_\odot$ model presented so far. 
For comparison, we have computed three other sets of stellar models with different convection descriptions: (i) with the Kuhfu\ss~1-equation model to compare the results of the 3-equation model to a simpler TCM; (ii) with MLT plus diffusive overshooting as described by \cite{freytag1996} to compare to one of the standard ad~hoc descriptions of convective overshooting with the same parameter value $f_\text{OV}=0.02$ as discussed above.  
Finally, (iii), we computed MLT models without overshooting to compare the results to a local convection theory. At least in terms of core size and temperature structure, models using the local Kuhfu\ss~theories would be equivalent to MLT models.
For the Kuhfu\ss~theory, we used the same value of 0.3 for the parameter $\alpha_\omega$, as before. To allow for a comparison across the mass range and the different convection descriptions, the models are selected at the same central hydrogen abundance of $X_{\rm c}=0.6$.

Figure~\ref{figmixedcorecompare} shows the sizes of mixed cores in models derived from these four descriptions of convection and convective overshooting over a range of initial masses. For all masses under consideration, the mixed core from the 3-equation model is larger than the convective core from an MLT model, as expected. This shows that when applying the 3-equation model, an overshooting zone emerges across the whole mass range investigated. Comparing the 3-equation model to the 1-equation model and the ad~hoc overshoot model, the mixed core sizes show good qualitative agreement. We repeat that this is achieved without fine-tuning any of the involved parameters. The relative size of the mixed cores decreases with decreasing stellar mass for all four descriptions, but differences in the details between the different convection descriptions are evident. For higher masses, the derived values for the mixed core sizes are almost identical among the ad~hoc overshooting and the 1- and 3-equation Kuhfu\ss~models. For low stellar masses, the results show larger discrepancies. Stellar models applying the 3-equation non-local model have the smallest cores, and the core size decreases faster with decreasing stellar mass than in the 1-equation models and the ad~hoc overshoot models. As the ad~hoc overshoot model has been calibrated to observations, this allows at least for an indirect comparison of the 1- and 3-equation model with observations.

In addition to the core sizes, we have also analysed the absolute values of the convective velocities. The Kuhfu\ss~model does not solve for the convective velocity itself, but rather for the TKE $\omega$. 
We approximate the mean convective velocity from the TKE by assuming full isotropy:
\begin{align}
v_\text{c,iso}=\sqrt{\frac{2}{3}\omega}\,.
\label{eqviso}
\end{align}
When using MLT, the convective velocities are computed as:
\begin{align*}
	v_c=\frac{1}{\sqrt{8}}\left(\frac{g}{H_p}\right)^{1/2}\Lambda\left(\nabla-\nabla_\mathrm{ad}\right)^{1/2}\,,
\end{align*}
\citep[e.g.][]{kippenhahn2012}. The computation of convective velocities is the same in MLT models with and without diffusive overshoot, as the inclusion of ad~hoc overshooting does not impact on the description of convection. To compare MLT and the Kuhfu\ss-theories we evaluate the maximum convective velocity in the convection zone. The maximum is reached well within the Schwarzschild boundaries for all models, and therefore allows for a consistent comparison.

\begin{figure}
	\centering
	\includegraphics[]{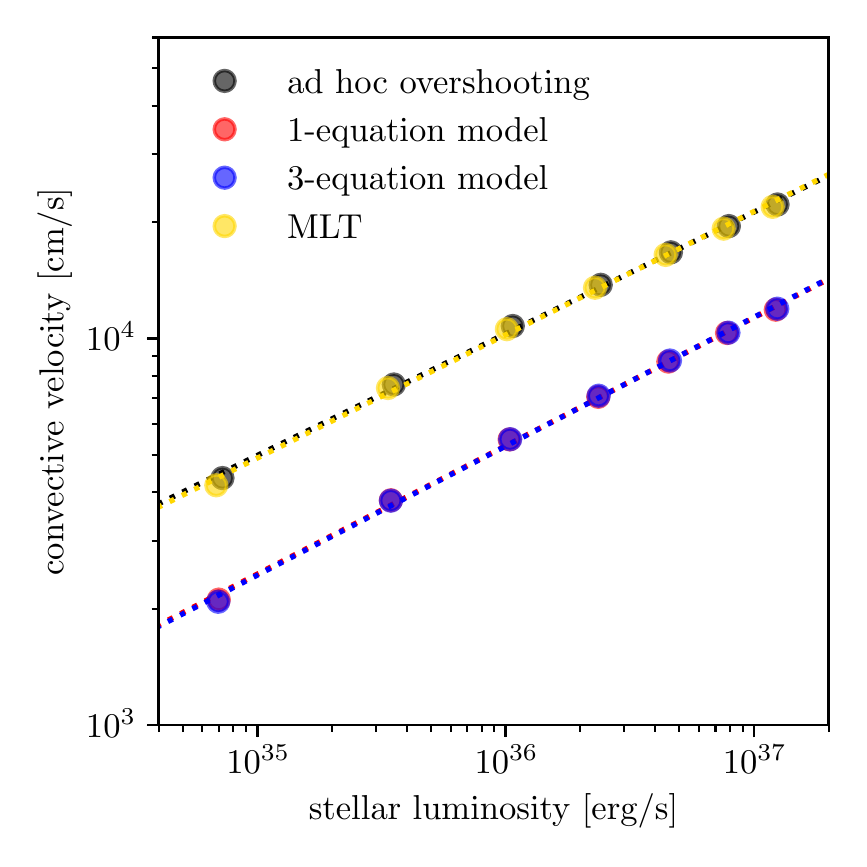}
	\caption{Comparison of the maximum convective velocities for different convection models as a function of stellar luminosity. For the Kuhfu\ss~model, the isotropic convective velocity $v_\text{c,iso}$ is plotted. The dotted lines indicate a linear fit to the logarithmic data. Please note that the data of the 1- and 3-equation Kuhfu\ss-models (red and blue points respectively) are largely overlapping, as well as the black and yellow dots.}
	\label{figenergyscaling}
\end{figure}

We show this comparison of the maximum convective velocities in the core as a function of stellar luminosity in Fig.~\ref{figenergyscaling}. The convection descriptions are the same as in Fig.~\ref{figmixedcorecompare}. For all cases, the scaling relation of the convective velocities has the same slope. A linear fit (dotted lines) to the data results in a value of $\sim0.3$ for all of them. However, the absolute values from the Kuhfu\ss~and MLT models differ by a constant factor of about two, indicated by the offset between the two pairs of lines. This difference in the absolute value is the result of two different effects. The change of the mixing length has the largest impact on the velocity. A reduced mixing length will lead to an increased dissipation rate and smaller velocities as a consequence. In the Kuhfu\ss~models we use a smaller value of $\alpha=1$ as obtained by a solar calibration instead of $\alpha=1.6$ for the MLT models. In addition, the mixing length is reduced towards the centre according to the \cite{wuchterl1995} formulation (see Eq.~\ref{eqdissold}). The convective velocity is reduced further compared to the local MLT models by taking the non-local terms into account, which act as a sink term in the bulk of the convection zone. The slope of 0.3 is very close to the $v_\text{c}\propto L^{1/3}$ scaling relation expected from MLT. This comparison also demonstrates that the absolute values of the TKE are very similar between the 1- and the 3-equation Kuhfu\ss~theories over the full mass range. For the $5\,M_\odot$ model, this was already apparent when comparing the TKE profiles in Fig.~\ref{fig1Eqcompare}.


%
%
\section{Discussion}
\label{sec:discussion}
\subsection{Relation between 1- and 3-equation model} \label{sepeclet}
As discussed in Sec.~\ref{sec1Eq} the 1-equation model is a simplification of the 3-equation model, for which \cite{kuhfuss1987} assumes that the convective flux is proportional to the super-adiabatic temperature gradient and the square-root of the TKE. This allows removing two of the three equations, namely for the entropy fluctuations $\Phi$ and the velocity entropy correlations $\Pi$. The equation for the TKE $\omega$ remains unchanged. The approximation for the convective flux allows expressing the temperature gradient as a function of the TKE. This couples the thermal structure and the TKE very closely. In the 3-equation model, the convective flux is evolved with an additional equation, which reduces the coupling of the thermal structure and the TKE. Despite the increased model complexity, the behaviour of the TKE in the 1- and 3-equation model is quite similar, as seen in Fig.~\ref{fig1Eqcompare}. 

In the bulk of the convection zone, both models result in the same absolute value of the TKE. This can be also seen in Fig.~\ref{figenergyscaling} by the agreement between red and blue points for a wider mass range. It can be attributed to the similarity of the equations. The additional dissipation is not or only weakly operating in the bulk of the convection zone. Hence, both the dissipation and the non-local term have the same functional form. The convective flux is adjusted such that a nearly adiabatic stratification is achieved in both models; therefore, the buoyancy term is also very comparable in the 1- and 3-equation models. As soon as the temperature gradient becomes subadiabatic, which happens already well within the Schwarzschild boundary in the 3-equation model, the dissipation by buoyancy waves is taken into account. This means that the functional form of the dissipation term changes. Due to the difference in the thermal structure in this region also the convective flux looks different. This will lead to a different solution for the TKE in the overshooting zone, as is evident from Fig.~\ref{fig1Eqcompare} towards the edge of the convection zone.

\begin{figure}
	\includegraphics{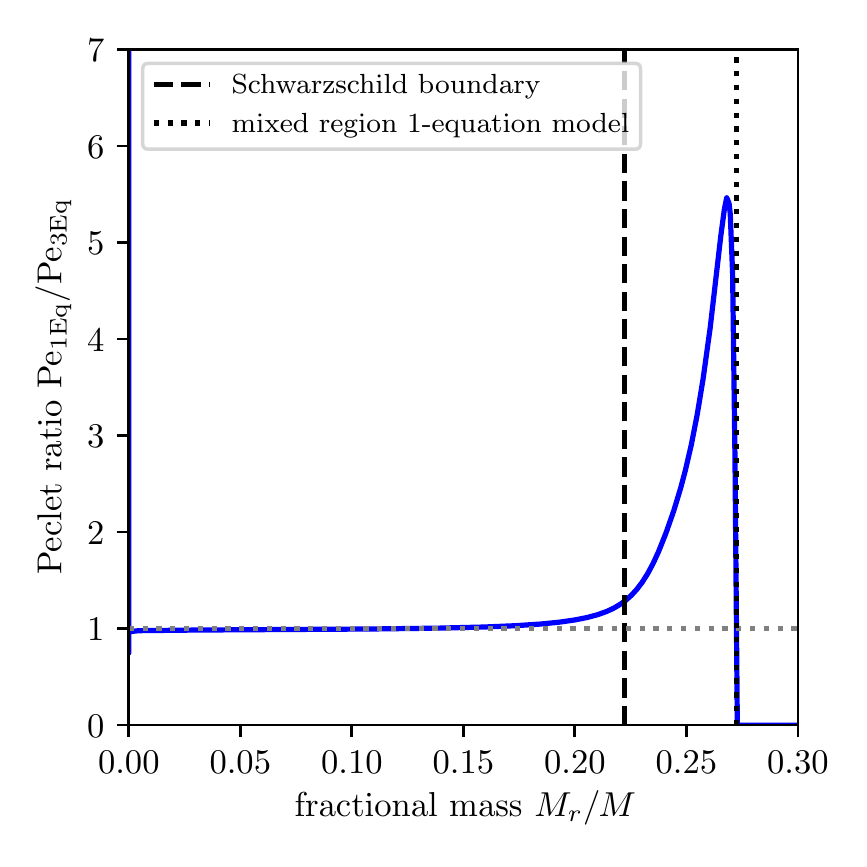}
	\caption{Ratio of the Peclet numbers for the 1- and 3-equation models for a $5\,M_\odot$ main-sequence model.}
	\label{figpecratio}
\end{figure}

The main difference between the 1- and 3-equation model is probably the temperature stratification 
which results from the solution of the model equations. Due to the coupling of the convective flux 
to superadiabatic gradient and convective velocities in the 1-equation model,
the convective flux is forced to have the same penetration depth 
into the stable layers as the TKE and at the same time to have the same sign as the superadiabatic gradient. 
As seen in our models and pointed out by \cite{xiong2001} this strong coupling leads to a nearly adiabatic temperature gradient 
in the overshooting zone and prevents the existence of a Deardorff layer (see also Paper~{\sc I}). Also with respect to other models 
of convection (e.g.\ MLT) the existence of a large subadiabatic zone in the convective region for the case of the 3-equation model is striking. To better understand 
the behaviour of the temperature gradient in the overshooting zone, we will analyse it in terms of the Peclet number. 
The Peclet number is the ratio of the timescales of radiative and advective transport and can be interpreted as an indicator 
for convective efficiency. A common definition of the Peclet number is:
\begin{align*}
\text{Pe}=\frac{u\cdot l}{\chi}
\end{align*}
where $u$ and $l$ are a typical velocity and length-scale of the convective flow. The radiative diffusivity is defined as
\begin{align*}
	\chi=\frac{16\sigma T^3}{3\kappa\rho^2c_p}\,.
\end{align*}
As a typical convective velocity we use again the isotropic velocity,  Eq.~(\ref{eqviso}). Due to the usage of typical scales for velocities and length-scales which are not rigorously defined, the interpretation of absolute values of the Peclet number remains difficult. Hence, we will only look at ratios of the Peclet number to compare different models. We will also assume that the length-scale and the radiative diffusivity are the same, when comparing different models. Under these assumptions, it is easy to see that:
\begin{align*}
	\frac{\text{Pe}_1}{\text{Pe}_3}\propto\sqrt{\frac{\omega_1}{\omega_3}}
\end{align*}
The ratio of the Peclet numbers obtained for the 1- and 3-equation models for a $5\,M_\odot$ main-sequence model is shown in Fig~\ref{figpecratio}. In the bulk of the convection zone, the 1- and 3-equation models have very similar Peclet numbers, which means the transport of energy by convection behaves very comparably. In the overshooting zone, however, the 1-equation model has a Peclet number which is up to 7 times higher than that of the 3-equation model. This indicates that convection as described by the 1-equation model is much more efficient in the overshooting zone than when described by the 3-equation model.

\begin{figure}
\includegraphics{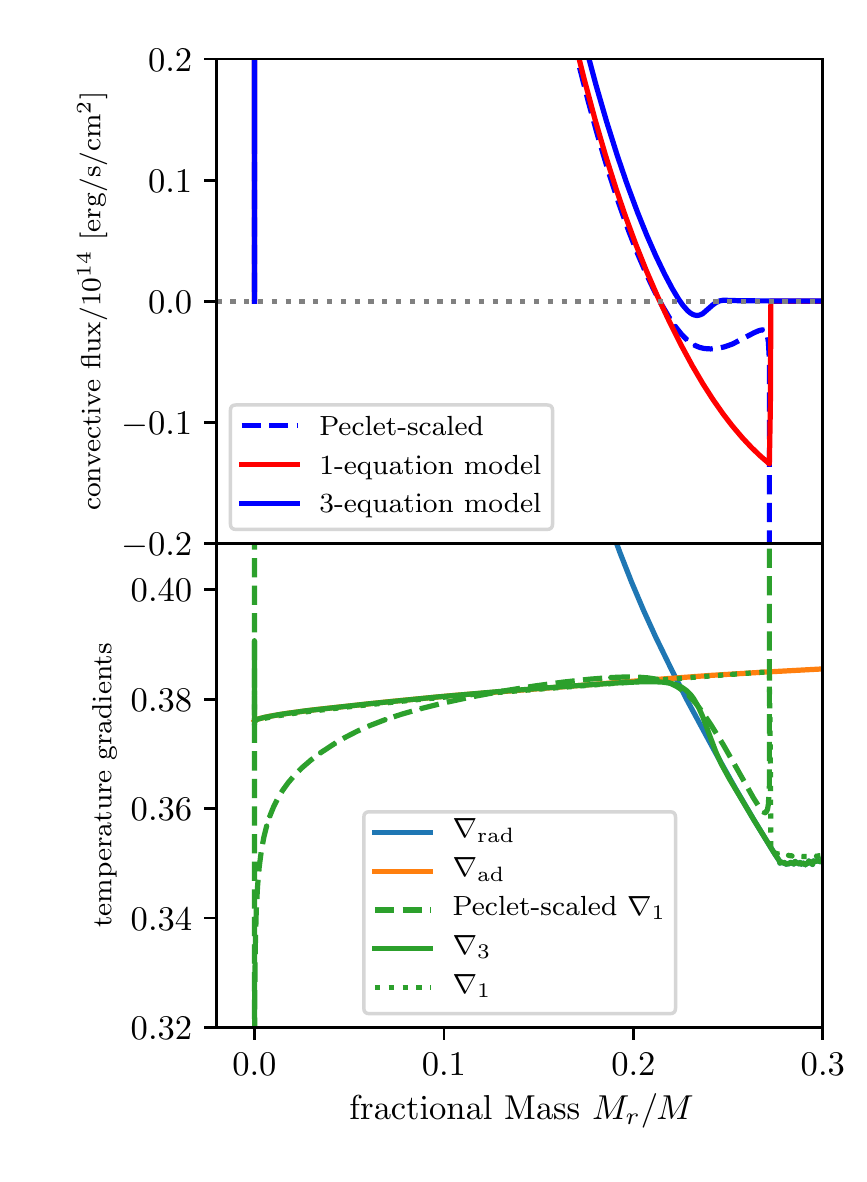}
\caption{\textit{Upper panel:} Convective flux as a function of fractional mass. The red and the blue lines show the results obtained from the 1- and 3-equation model, respectively. The blue dashed line indicates the convective flux scaled with the Peclet number according to Eq.~(\ref{eqfluxPe}). \textit{Lower panel:} Temperature gradients as a function of fractional mass. The blue and orange lines indicate the radiative and adiabatic temperature gradient, respectively. The green dotted and solid lines indicate the temperature gradient as obtained by the 1- and 3-equation models. The green dashed line shows the temperature gradient of the 1-equation model computed from the scaled convective flux shown in the upper panel.}
\label{figpecgrad}
\end{figure}

This change in efficiency will also impact the temperature gradient. Following the approximation of the convective flux in the 1-equation model, the convective flux scales as $\Pi\propto\sqrt{\omega}$. Using the ratio of the Peclet numbers, one can therefore write a Peclet-scaled convective flux
\begin{align}
\Pi_{3,\text{Pe}}=\Pi_1\cdot\frac{\text{Pe}_3}{\text{Pe}_1},
\label{eqfluxPe}
\end{align}
to mimic the convective flux in the 3-equation model. The resulting convective flux is shown with a blue dashed line 
in the upper panel of Fig.~\ref{figpecgrad}. Using this scaled convective flux, a scaled temperature gradient can be computed, 
as illustrated by the green dashed line in the lower panel of Fig.~\ref{figpecgrad}. For comparison, the model temperature gradients 
of the 1- and 3-equation model are visualized by a green dotted and solid line, respectively. 

This comparison confirms that the reduced convective efficiency obtained from the Peclet numbers is sufficient to change the behaviour of the temperature gradient from nearly adiabatic to more radiative in the overshooting zone, and implies that the behaviour of the temperature gradient in the overshooting zone can at least qualitatively be predicted from the TKE alone without invoking the other convective equations for $\Pi$ and $\Phi$. The fact that the region of negative values in the scaled convective flux is more extended compared to the actual convective flux from the 3-equation model is due to the different penetration depths of TKE and convective flux in the 3-equation model. We conclude that this is an important indication for internal consistency of the model. It furthermore shows that the mostly radiative temperature gradient in the overshooting zone is in fact a result of reduced convective efficiency in the overshooting zone in the 3-equation model. Following the terminology proposed by \cite{zahn1991}, in the 1-equation model the overshooting zone is best described by subadiabatic penetration while the more inefficient convection in the 3-equation model concerns overshooting of chemical element distributions only.

Considering the chemical mixing, both models will result in a more step-like chemical mixing profile. 
The extent of the mixed region is mainly dependent on the choice of the parameter $\alpha_\omega$, 
as discussed in Appendix~\ref{secparamdep}. This parameter cannot be determined from first 
principles. As for the ad hoc descriptions of convective core overshooting, a calibration of this parameter is required, 
which will be discussed below. Furthermore, the resulting convective flux is also very similar in the bulk of 
the convection zone and differences become only obvious in the overshooting zone. Given the relative freedom 
in choosing $\alpha_\omega$ and the similarity in the mixing properties of both models, resulting stellar models 
are basically indistinguishable when comparing the chemical structure. Once the observations become sensitive 
enough to the thermal structure in the overshooting zone as, for example, with the help of asteroseismology
\citep{michielsen2019}, it will become possible to detect
differences between both models. In view of the general agreement 
between 1- and 3-equation model, the application of the 1-equation model seems to be sufficient to obtain the chemical structure 
from a non-local convection model. The parameter $\alpha_\omega$ can be tuned to obtain the correct size of the convective 
core. However, the stratification obtained from the 1-equation model is less realistic.

%
\subsection{Other constraints on core overshooting}
To date, a range of different approaches to determine the extent of convective cores has been followed. In this subsection, we discuss a few examples of convective core size determinations. The need for a larger mixed core has been recognized already in the '80s of the last century (see, for example, \citealt{bressan1981} in relation to the Hertzsprung-Russell-diagram of massive stars, or \citealt{maeder1981} concerning that of open clusters). In the latter case, isochrones derived from stellar models which include core overshooting match the morphology of the turn-off region better than models without overshooting \citep{pietrinferni2004,magic2010}. The comparison to observations showed further that the overshooting distance in terms of pressure scale heights needs to increase with mass in the range between 1.2 and $\sim2\,M_\odot$. This mass dependence can be included explicitly in the computations by expressing the overshooting parameter as a function of the total stellar mass. Alternatively, this mass dependence can also be introduced in the stellar models by limiting the radial extent of the overshooting zone geometrically. In both cases, the parameter of the overshooting scheme effectively needs to increase with stellar mass. For the TCM, however, this is a natural outcome  without imposing it.

Eclipsing binary systems offer an excellent opportunity to put constraints on stellar physics. \cite{claret2019} used a large sample of eclipsing binaries to determine the overshooting parameter as a function of mass. They find a clear increase of the overshooting parameter (extent) with mass, even though the statistical significance of this result has been debated \citep{constantino2018}. In a detailed analysis, \cite{higl2018} addressed the evolution of the binary system TZ~Fornacis with an evolved red giant primary and a main-sequence secondary star, both with masses of $\sim2\,M_\odot$. They found that a basically unrestricted overshooting extent, using the standard value for the free parameter, is required to explain the evolution of this system. This puts further constraints on the mass dependence of the overshooting parameter. 

The high precision photometric data obtained from space telescopes like \textit{Kepler} or CoRoT allowed setting
further constraints on the stellar evolution models. Using asteroseismology of g-mode pulsators, the convective 
core masses of intermediate mass stars has been determined for larger samples of stars \citep[e.g.][]{pedersen2021,mombarg2019}. 
Similarly, the seismology of p-mode pulsators allows determining the required overshooting efficiency in lower mass stars 
\citep[e.g.][]{deheuvels2016,angelou2020}. In agreement with the previously mentioned studies, they find that the relative 
mass of the mixed core needs to increase with stellar mass. A more detailed comparison of the TCM models discussed in this 
work with asteroseismic observations needs to be addressed in future work. Finally, asteroseismology allows probing the 
temperature gradient in the overshooting zone. \cite{michielsen2021} inferred a predominantly radiative overshooting zone 
in a $\sim3.5\,M_\odot$ main-sequence star, which is in agreement with the temperature gradient obtained from the 3-equation 
model but disagrees with the 1-equation model. However, they point out that this result is only obtained for this single B-type star and might not be generalisable for all B-type stars.

With the increase of computational resources in recent years, more and more  multidimensional hydro-simulations of stellar 
core convection have been carried out \citep[e.g.\!][]{meakin2007,gilet2013,edelmann2019,higl2021}. These simulations 
confirmed, for example, the scaling of the stellar luminosity with the third power of the convective velocities 
\citep[e.g.][and references therein]{edelmann2019,higl2021}. In \cite{higl2021} the authors calibrate the overshooting 
parameter $f_\text{OV}$ in GARSTEC to 2D simulations of core convection in low and intermediate mass stars. By matching 
the size of the mixed convective core in the 1D GARSTEC models to the size of the mixed region in the 2D simulations, they 
find that the effective parameter of $f_\text{OV}$ needs to decrease with stellar mass. To limit the size of the convective 
cores for small stellar mass in the 1D models, they use the geometric ``square'' cut-off according to Eq.~(\ref{eqcut2}). 
As they use the same stellar evolution code as we do, this allows for a direct comparison of the results. \cite{higl2021} 
find that the size of the convective cores resulting from the 2D simulations need to be larger than the GARSTEC models 
computed including the square cut-off (Eq.~\ref{eqcut2}) at a constant overshooting parameter. This indicates that the 
geometric square cut-off is too restrictive. As the 3-equation model predicts mixed core masses similar to the GARSTEC 
models including this geometric cut-off (see Fig.~\ref{figcompareALL}) this indicates that our 3-equation model might 
be too restrictive as well at this lower mass range. 

Finally, other TCMs have been used to compute stellar evolution models including the effects of non-local convection. 
\cite{xiong1986} has computed stellar models in the mass range of 7 to $60\,M_\odot$. He finds that by solving the 
convection equations, the TKE extends beyond the formal Schwarzschild boundary and that this also increases the 
size of the mixed region. The size of the well mixed region increases with increasing stellar mass. Both results 
are in agreement with our findings employing the 3-equation model. \cite{xiong1986} also finds that the convective 
flux penetrates much less deeply into the stable layers than the TKE. The magnitude of the convective flux in 
this region is negligible, which causes the temperature gradient to be mostly radiative in the overshooting 
region. This is in good agreement with the decoupling of the thermal and the chemical structure we discussed 
in detail in Sect.~\ref{sec:models} and Paper~{\sc I}. \cite{zhang2016} applied the TCM by \cite{li2007} in a 
similar mass range as in this work. They develop a simplified model comparable to the 1-equation model 
and find very good agreement between the full model and the simplified version. \cite{li2017} applies the 
simplified TCM by \cite{li2012} to compute stellar models of a $5\,M_\odot$ star and find an overshooting 
distance of about $0.2H_p$, which is comparable to the overshooting distance obtained with our 3-equation model.
%
%
\section{Conclusions}  \label{sec:conclusions}
In this work, we have presented results of stellar structure and evolution calculations using the TCM proposed by \cite{kuhfuss1987}. 
We have implemented the Kuhfu\ss~model into GARSTEC which solves the four stellar structure equations and the three equations of the 
convection model simultaneously with the implicit Henyey method. We have computed main-sequence models of intermediate-mass main-sequence stars 
between $1.5$ and $8\,M_\odot$ which consistently compute the structure and evolution of the TKE, convective flux, and 
entropy fluctuations. This naturally includes the effects of convective overshooting for the thermal and chemical structure. In 
Paper~{\sc I}, we have demonstrated that the original 3-equation model with standard MLT prescription for the dissipation length 
of TKE leads to convection zones which essentially extend throughout the entire stellar interior. We have therefore implemented the 
dissipation by gravity waves as discussed in Paper~{\sc I} in addition to the original Kuhfu\ss~model. We showed that the Kuhfu\ss~3-equation 
model with an increased dissipation rate results in models with physically reasonable overshooting distances. This indicates that the 
dissipation was actually underestimated by the original description, and that dissipation by gravity waves is a relevant effect in core 
overshooting zones.

In Fig.~\ref{figsummary} here we have shown a summary of the TKE and flux and the temperature gradient for a convective 
core of a $5\,M_\odot$ main-sequence model. We find that the TKE extends beyond the formal Schwarzschild boundary of 
convective neutrality. This is the result of the non-local terms in the Kuhfu\ss~model. The convective flux shows a region of negative 
values beyond the Schwarzschild boundary, which is, however, penetrating less deeply into the stable layers than the TKE. As the 
convective motions are very efficient in mixing chemical elements, the extended convective core will have essentially the same 
composition as the convective core. This can be seen in Fig.~\ref{figcompareMLT}, upper panel, in which the hydrogen profiles at 
the end of the main-sequence of an MLT and a Kuhfu\ss~3-equation model are compared. In the Kuhfu\ss~model, this extension beyond 
the Schwarzschild boundary is the outcome of the solution of the model equations and not due to the inclusion of any sort of ad~hoc 
overshooting. We also compared the results of the full 3-equation model to the simplified 1-equation model and find qualitative 
and quantitative agreement of the TKE throughout a large part of the convection zone. This is a result of the similarity 
of the model equations and the chosen parameters, which are the same for both models. In addition to the convective velocity and 
the associated mixing also the temperature gradient is part of the model solution. This is another important difference compared 
to ad~hoc descriptions of convective overshooting, in which the temperature gradient needs to be assumed separately and independently.

The analysis of the temperature gradient has shown the existence of a Deardorff-layer \citep{deardorff1966}, 
in which the temperature gradient is subadiabatic and the convective flux is still positive. The existence of the Deardorff-layer 
has been confirmed in different numerical simulations of stellar convection \citep{chan1992,muthsam1995,muthsam1999,tremblay2015,kapyla2017} 
and other Reynolds stress models \citep{kupka1999b,xiong2001,kupka2002,montgomery2004,zhang2012a}. 
Beyond the Deardorff-layer, the convective flux becomes negative as the result of the stable stratification 
(cf.\ also \citealt{muthsam1995} and references in \citealt{canuto1992}). In the overshooting region, the model temperature gradient 
gradually transitions from a slightly subadiabatic to a radiative value, exhibiting a small region with a super-radiative temperature gradient 
(see Fig.~\ref{fignabla}). However, this transition region is rather narrow, such that the overshooting zone has a mostly radiative temperature 
gradient. This is in agreement with very recent results from asteroseismology \citep{michielsen2021}. In contrast to the 3-equation model, 
the overshooting zone of the 1-equation model shows a mostly adiabatic temperature gradient. 
As pointed out by \cite{xiong2001} this can be attributed to the assumption of a full correlation between the convective flux 
and the convective velocities, as it is done in the 1-equation model (see Eq.~\ref{eqconvflux1}). The approximation 
Eq.~(\ref{eqconvflux1}) does not allow for a Deardorff layer, as the convective flux is proportional to the superadiabatic gradient. 
From a theoretical point of view, the existence of a Deardorff-layer can be attributed to the non-local term of the 
$\Phi$-equation Eq.~(\ref{eqKuh3}) \citep{deardorff1966}. This shows that an independent equation for the convective flux 
is required \citep[see also][]{kupka2020} and highlights the necessity to consider
more complex turbulence models like the 3-equation model to capture the temperature structure in the overshooting zone 
more accurately. The comparison of the Peclet numbers of the 1- and 3-equation models shows further that the mostly 
radiative temperature gradient in the overshooting zone of the 3-equation model can be explained by the reduced 
TKE/velocities compared to the 1-equation model. The narrow range of the transition region from 
an adiabatic to a radiative temperature gradient can be attributed to the shallower penetration of the convective flux into 
the stable layers compared to the 1-equation model. 

A comparison with stellar models using MLT shows qualitative agreement of the TKE and the convective flux in the part of the convection zone unstable according to the Schwarzschild criterion ($\nabla_{\rm ad}=\nabla_{\rm rad}$). For the convective flux, we even find a very good quantitative agreement between the Kuhfu\ss~model and MLT in that region (see Fig.~\ref{figcompareMLT}, lower panel). The convective velocities found in the Kuhfu\ss~model are smaller than in MLT by a factor of two. This qualitative agreement indicates that the stellar structure has the largest impact on the convective properties in the bulk of the convection zone, irrespectively of the convection model in use. The nuclear energy released in the centre determines the convective flux --- about 80\% of the local flux in the centre --- and the coefficients of the convection model determine the absolute values of the other convective variables. Differences appear in the overshooting zone, which is sensitive to more subtle changes in the convection model. In the overshooting zone the stellar flux is mainly transported by radiation, such that the convective structure of this region is less constrained by the stellar structure but a result of the convection model.

The results of the 1- and 3-equation models over a broader mass range show qualitative agreement with other overshoot 
descriptions. For given values of $\alpha_\omega$ and $f_{\rm OV}$, this is achieved without fine-tuning any of the other model parameters, as we have used the closure parameters 
suggested by the authors of the turbulence model \citep[see ][ and references therein]{canuto1998}. Tests of those 
parameters --- for different physical scenarios --- are published in the literature \citep{kuhfuss1986,kuhfuss1987,wuchterl1995,canuto1992,canuto1998,kupka2007c}. 
In Fig.~\ref{figmixedcorecompare} we have shown a comparison of mixed hydrogen core sizes computed with diffusive 
overshooting, the 1-equation and the 3-equation model, and MLT. The comparison shows that the exact extent of the 
convective core depends on the details of the model. For the same parameter choice of the parameter $\alpha_\omega$ 
at lower masses, the 3-equation model shows a reduced amount of overshooting compared to the 1-equation model, 
while the 3-equation models have larger cores at higher masses. The newly introduced parameters which control the 
reduction of the dissipation length scale $\Lambda$ were shown to have a moderate impact on the overshooting extent. 
For its default overshoot parameter, the diffusive mixing model produces yet larger mixed core sizes. Towards the lower 
end of the mass range, both Kuhfu\ss~models show a decrease of the mixed core size as implemented by other methods 
to match observations from open clusters and binaries \citep{claret2019,pietrinferni2004,magic2010}. \cite{higl2018} 
found that the overshooting parameter needs to increase steeper with mass than predicted by GARSTEC including the 
geometric square cut-off from the analysis of the TZ~For binary system. In agreement with the conclusion from the 2D 
simulations, \citep{higl2021} this indicates that the convective core size predicted by the 1- and 3-equation model 
increases too shallowly with stellar mass for a constant parameter $\alpha_\omega$. To match the observations and the 
results from the 2D simulations, this would require to modify the parameter $\alpha_\omega$ as a function of mass. 
Further constraints on the parameter $\alpha_\omega$ can be obtained by comparing the Kuhfu\ss~model to results from 
asteroseismology of intermediate mass stars.

However, such tuning of $\alpha_\omega$ seems non-advisable: the physical incompleteness of the ad~hoc model 
of convective overshooting, as an example, is well demonstrated by the shrinking of the convective core size with 
mass for stars with $M<2\,M_\odot$ not only requiring an extra cut-off function (Eq.~\ref{eqcut2}) to limit 
the convective core size to values compatible with observations, but even demanding a more fine-tuned function 
(Eq.~\ref{eqcuttanh}) to pass such a stringent test. While applying a similar procedure to $\alpha_\omega$ as 
for $f_\text{OV}$ appears to be a convenient ad~hoc solution to match exactly the observational data, it provides 
no new insights and requires redoing similar procedures in related, but different scenarios. Finding the physical 
reason for the remaining, now already much smaller discrepancies with the data, on the other hand, might allow for a 
model not requiring such measures in other applications either (cf. also the discussion on such requirements in 
\citealt{kupka2017}).

Our study of stellar models applying the 3-equation model from \cite{kuhfuss1987} has shown that the resulting 
stellar structure depends sensitively on the details of the convection model. Although the original 1- and 3-equation 
theories are very closely related, their application results in very different structures of the convection zone (see Paper~{\sc I}). 
We demonstrated that modifying the dissipation term of the TKE can remedy this discrepancy. 
The improved 3-equation model compares very well with the 1-equation model, both in terms of the TKE 
and the mixing properties. For applications in which the temperature structure of the overshooting zone is not important, 
the 1-equation model describes the convective core similarly well as the 3-equation model. The parameter $\alpha_\omega$ 
allows obtaining the required convective core size to match with the observations. Only when the thermal structure of the 
overshooting zone is of major interest, a more complex convection model like the 3-equation model needs to be used. 
A self-consistent prediction of the detailed convective core structure appears to require a physically more complete 
(and thus more complex model). As discussed above, at least three equations are required to allow for more complex 
phenomena like, for example, a Deardorff layer. In \cite{canuto1998} further partial differential equations for the dissipation rate 
and the vertical component of the TKE are discussed. This potentially allows removing some of the simplifications 
still present in the 3-equation model. However, increasing the number of equations comes at the price of gaining numerical complexity. 
Including these equations into stellar structure and evolution models has to remain the task of future work.
Finally, a comparison with more and more realistic 3D hydro-simulations of convective cores will be useful 
to put further constraints on our TCM model, in particular to restrict closure conditions. This is presently under way 
(Ahlborn \& Higl, in preparation). The application of the extended 3-equation Kuhfu{\ss} model to stellar envelopes 
will be presented in another forthcoming paper (Ahlborn et al.~2022, in preparation). In any case, turbulent convection 
models offer a convincing and feasible improvement of the treatment of convection in 1D stellar models beyond 
the standard mixing-length theory.

\begin{acknowledgements}
FK is grateful to the Austrian Science Fund FWF for
support through projects P29172-N and P33140-N and support from
European Research Council (ERC) Synergy Grant WHOLESUN
\#810218
\end{acknowledgements}
\bibliographystyle{bibtex/aa} 
\bibliography{hydro}
\appendix
\section{Comparison of 1- and 3-equation model}
\begin{figure}[!htb]
\centering
 \includegraphics[width=\columnwidth]{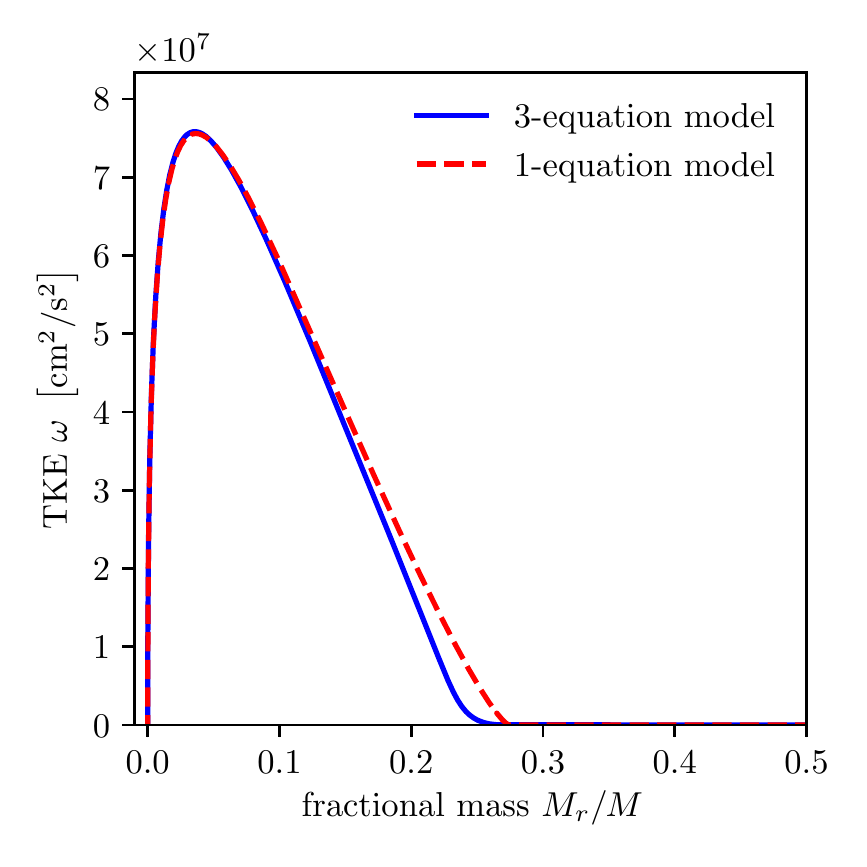}
 \caption{Comparison of the TKE of the 1-equation and 3-equation non-local models in a $5\,M_\odot$ main-sequence model 
              with limited dissipation length scale $\Lambda$
for the 3-equation model.}
 \label{fig1Eqcompareuntrans}
\end{figure}

Figure~\ref{fig1Eqcompareuntrans} shows the TKE in the 1- and 3-equation model on a linear scale. Even though the TKE of the 3-equation model has approximately the same extent as the 1-equation model as seen in Fig.~\ref{fig1Eqcompare} it drops off faster. Despite the smaller values of the TKE the efficiency of the chemical mixing is still high enough to fully mix the overshooting region. Therefore, the chemical profiles obtained from the 1- and 3-equation model are very comparable.

\section{Parameter dependence}
\label{secparamdep}
\begin{figure}
\centering
 \includegraphics[width=\columnwidth]{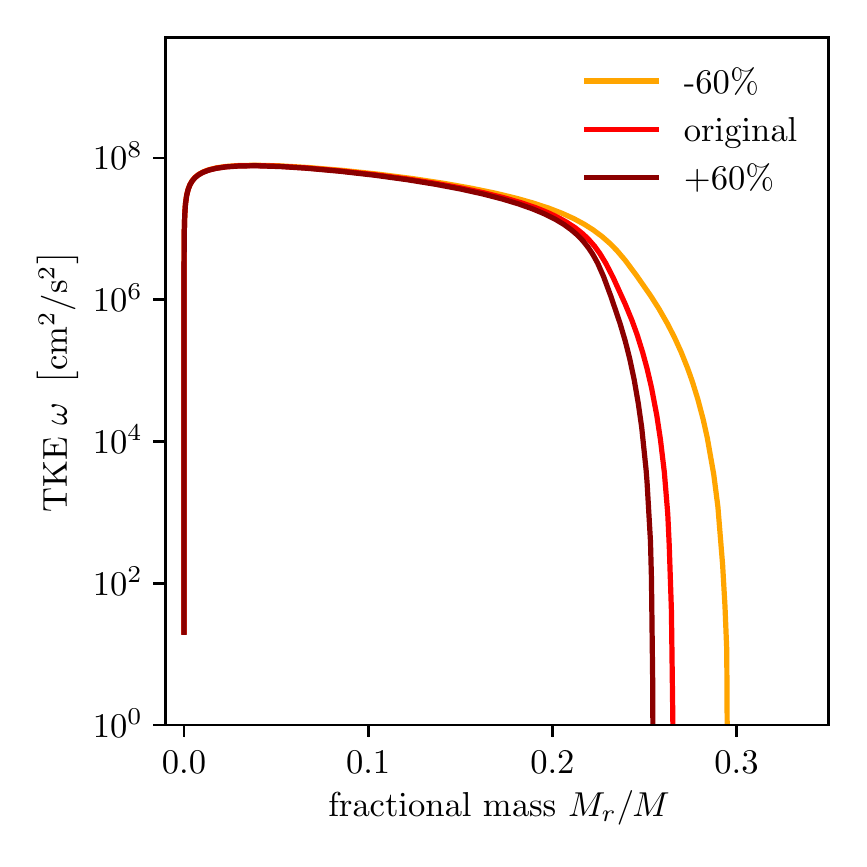}
 \caption{Comparison of the TKE as a function of fractional mass on a logarithmic scale for different values for the parameter $c_4$ in a 3-equation non-local $5\,M_\odot$ main-sequence model 
 with limited dissipation length scale $\Lambda$.}
 \label{figvariation}
\end{figure}

\begin{figure}
\centering
 \includegraphics[width=\columnwidth]{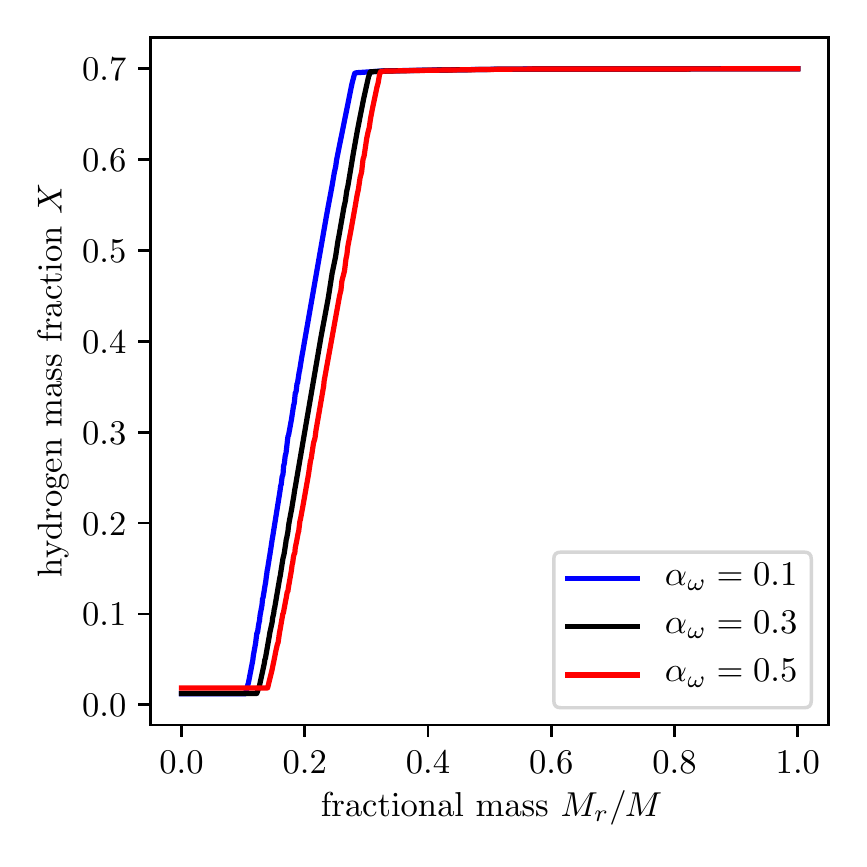}
 \caption{Comparison of the hydrogen profiles as a function of fractional mass for different values for the parameter $\alpha_\omega$ in a 3-equation non-local $5\,M_\odot$ main-sequence model 
             with limited dissipation length scale $\Lambda$.}
 \label{figovcomparison}
\end{figure}

The model for the modification of the dissipation length scale comes with a number of parameters ($c_2,c_3,c_\epsilon$). 
These parameters are not necessarily free, but have been calibrated in the framework of the convection models developed 
by \cite{canuto1992}. These parameters enter the equation of the reduction factor as a single parameter, which was previously 
defined as $c_4$. Given the dissipation parameter of the 3-equation model, we find $c_4\approx0.072$ 
while for the dissipation parameter of the \citet{canuto1998} model $c_4\approx0.2$ as described in Paper~I. 
As mentioned already in Sect.~3.6 in Paper~I, the effect of changing $c_\epsilon$ on changing $c_4$ to some extent 
cancels out in the calculation of $\Lambda$, as $c_\epsilon$ appears in the denominator of $c_4=c_3/(c_2 c_{\epsilon})$ and in the
numerator of $\Lambda = c_{\epsilon} \omega^{3/2} / \epsilon$. Hence, if $c_4$ is adjusted according to $c_4=c_3/(c_2 c_{\epsilon})$,
a change of $c_\epsilon$ first of all influences the TKE dissipation rate $\epsilon$ throughout the whole model and is
not specifically changing $\epsilon$ only within the overshooting zone.
To check which impact these parameters actually have on the result, we varied their values. 
As they enter the equations as one parameter, we only varied this effective parameter value $c_4$ by $\pm 60\%$. 
All other parameters take their default values. The resulting profiles of the TKE are shown in Fig~\ref{figvariation}. 
It can be seen that the variation of the parameter $c_4$ leads to some noticeable variation in the TKE profile. 
However, within these ranges, the models keep their property of a limited overshooting range. 
The direction of the variation can be explained by the theory as well. An increase of the parameter will lead to a decrease 
of the dissipation length scale $\Lambda$. A decreased length scale leads in turn to an increased dissipation. 
This reduces the overshooting which can be observed for the dark red line. For the case with a decreased parameter value, 
the same argument applies in the opposite direction. The expected behaviour can be similarly observed for the yellow line.\\
Apart from the new parameters for the reduction of the dissipation length scale $\Lambda$, the model still contains 
the original parameters of the Kuhfu\ss~model. Here, we focus on the parameter $\alpha_\omega$ 
which controls the flux of the TKE. In the 1-equation non-local model, this parameter controls 
the extent of the overshooting region. Kuhfu\ss~suggested a default value of $\alpha_\omega=0.25$. 
In Fig.~\ref{figovcomparison} we show three different hydrogen profiles for the $5\,M_\odot$ model at the same age. 
The parameter $\alpha_\omega$ takes values of 0.1, 0.3 and 0.5. 
It can be clearly seen that the original property of this parameter of controlling the overshooting extent is still given.
\begin{figure}
	\centering
	\includegraphics[width=\columnwidth]{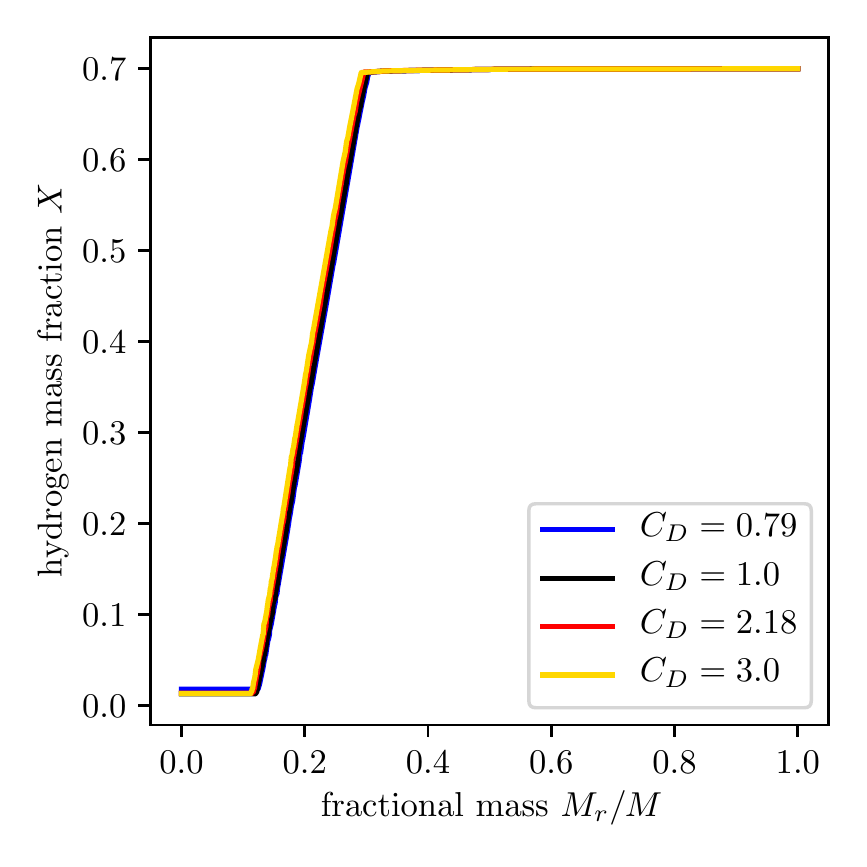}
	\caption{Comparison of the hydrogen profiles as a function of fractional mass for different values for the parameter $C_\text{D}$ in a 3-equation non-local $5\,M_\odot$ main-sequence model with limited dissipation length scale $\Lambda$.}
	\label{figovcomparisonCD}
\end{figure}
\begin{figure}
	\centering
	\includegraphics[width=\columnwidth]{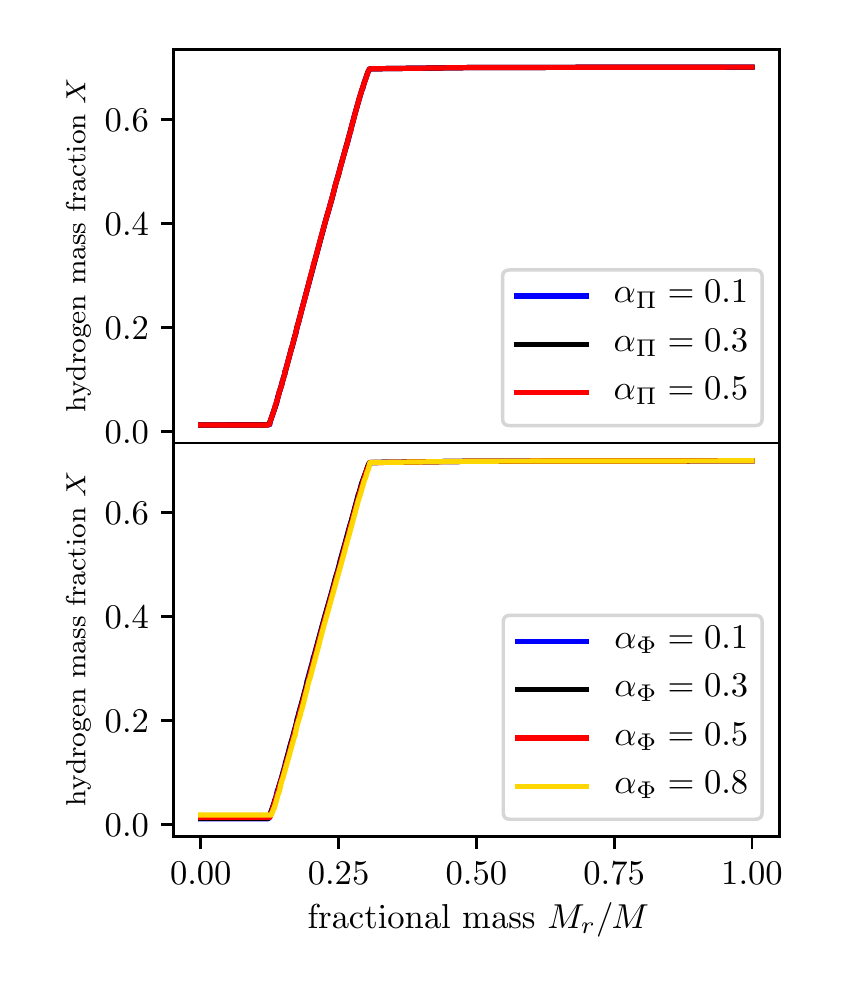}
	\caption{Comparison of the hydrogen profiles as a function of fractional mass for different values for the parameters $\alpha_\Pi$ and $\alpha_\Phi$ in a 3-equation non-local $5\,M_\odot$ main-sequence model 
		with limited dissipation length scale $\Lambda$.}
	\label{figovcomparisonPiPhi}
\end{figure}
In Fig.~\ref{figovcomparisonCD} we show hydrogen profiles of a $5\,M_\odot$ star at the end of the main-sequence 
for different values of the dissipation parameter $C_\text{D}=0.79,1,2.18$ and 3. 
We always assume $C_\text{D}=c_{\epsilon}$, as both parameters have the same role in the 
\citet{canuto1998} and in the Kuhfu\ss~model. Here, the value of 0.79 refers 
to the default dissipation parameter from the \citet{canuto1998} model, while 2.18 is the numerical default value in the Kuhfu\ss~model. The extent of the hydrogen 
profile is largest for the model computed with the smallest dissipation parameter and smallest 
for the largest parameter. This behaviour is expected, as a decreased dissipation allows the 
TKE flux to extend further out. Compared to the parameter of the TKE 
flux $\alpha_\omega$, the impact of the dissipation parameter on the overshooting extent is 
rather limited, as the variation is much smaller when compared to the results shown in 
Fig.~\ref{figovcomparison}. The variation of the overshooting extent is also smaller, as one 
could have expected from the comparison shown in Fig.~\ref{figvariation}. This is because by 
changing $C_\text{D}$ also $c_4$ will change, while in Fig.~\ref{figvariation} only $c_4$ is 
changed. The effects of changing $c_4$ and $C_\text{D}$ partially compensate each other, resulting 
in a smaller net effect. Finally, one could find combinations of parameters $\alpha_\omega$ 
and $C_\text{D}$ which allow obtaining models with equal convective core sizes. In 
Fig.~\ref{figovcomparisonPiPhi} we show the impact of the parameter $\alpha_\Pi$ and $\alpha_\Phi$ in 
the upper and lower panel, respectively. They both have a negligible impact on the overshooting distance, 
which is yet smaller than the impact of the dissipation parameter $C_D$. This indicates again that the 
parameter $\alpha_\omega$ which determines the importance of the TKE flux has the 
largest impact on the overshooting distance (note that Fig.~\ref{figvariation} has a different scale for the fractional mass axis).

\section{Ad hoc overshooting for small convective cores}
\label{seccut}
\begin{figure}
	\centering
	\includegraphics{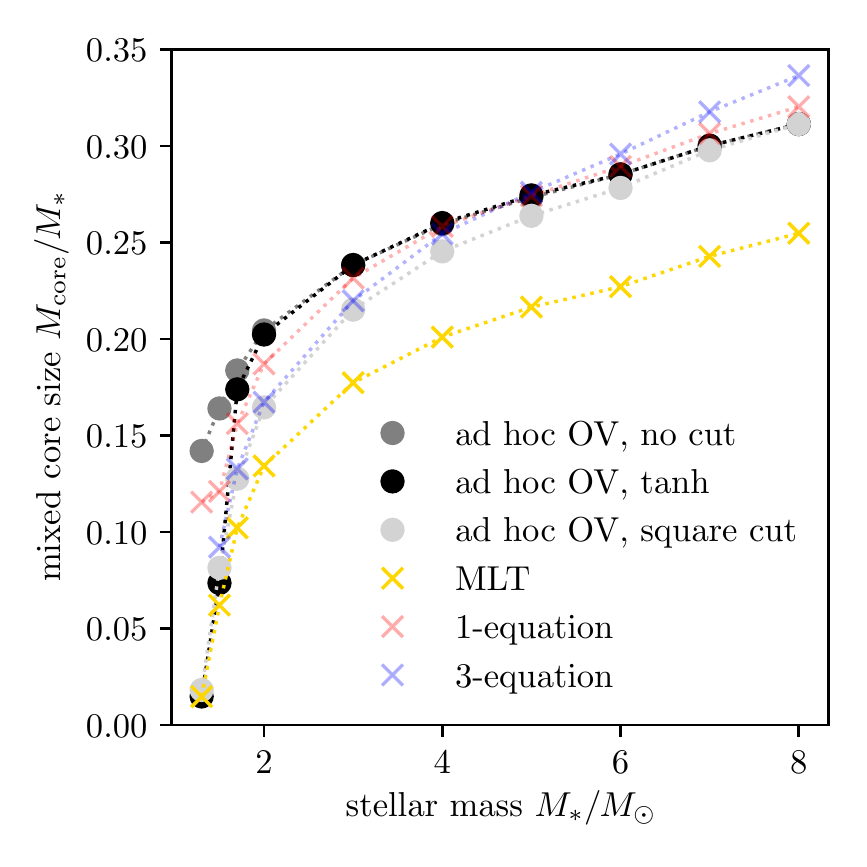}
	\caption{Comparison of mixed core sizes obtained with different descriptions of convection and different geometric cut to limit the overshooting (OV) for small cores in the ad~hoc overshooting model.}
	\label{figcompareALL}
\end{figure}
Parametrised descriptions of overshooting like the diffusive or step overshooting use the pressure scale-height at the Schwarzschild boundary to define the overshooting distance. In the description of \cite{freytag1996} the diffusion coefficient is computed according to:
\begin{align}
D_{\rm adhoc}=D_0\exp\left(\frac{-2z}{f_{\rm OV}H_p}\right)
\label{eqdiffcoeffadhoc}
\end{align}
where $f_{\rm OV}$ is an adjustable parameter and $z$ is the radial distance to the Schwarzschild boundary.
The pressure scale-height is, however, diverging towards the stellar centre. Hence, the inferred overshooting distance for a fixed overshooting parameter will increase for a decreasing convective size as well as the size of the mixed region. The resulting mixed core sizes for a fixed parameter value have been shown to be too large when comparing to observations \citep{pietrinferni2004,magic2010}. To avoid the unphysical growth of the overshooting region, the overshooting parameter needs to be artificially restricted. In GARSTEC the unphysical growth of the overshooting zone is prevented by comparing the size of the Schwarzschild core to the pressure scale height and use the smaller one as the relevant length-scale. As usual, there are different ways to implement this. Originally, \cite{magic2010} suggested the following expression:
\begin{align*}
	\widetilde{H_p}=H_p\cdot\mathrm{min}\left(1,\left(\frac{r_\mathrm{CZ}}{H_p}\right)^2\right)
\end{align*}
where $r_\mathrm{CZ}$ is the radius of the Schwarzschild boundary. A correction factor of the same type was also used in \cite{higl2021} when comparing GARSTEC results to 2D simulations, while introducing a factor of 2 in the denominator:
\begin{align}
\widetilde{H_p}=H_p\cdot\mathrm{min}\left(1,\left(\frac{r_\mathrm{CZ}}{2H_p}\right)^2\right)
\label{eqcut2}
\end{align}
As the size of the convective core is now compared to a length scale twice as large as the original expression the size of the overshooting region is limited more strongly by the latter expression. The comparison to 2D simulations \citep{higl2021} as well as the study of the eclipsing binary TZ~For \citep{higl2018} revealed, however, that this expression is finally limiting the size of the convective core too strongly. This led to the introduction of a different functional form of the limitation:
\begin{align}
\widetilde{H_p}=H_p\cdot\mathrm{min}\left(1,\frac{1}{2}\left[\mathrm{tanh}\left(5\left(\frac{r_\mathrm{CZ}}{H_p}-1\right)\right)+1\right]\right)
\label{eqcuttanh}
\end{align}
which is less strongly limiting the core sizes at $2\,M_\odot$ but very quickly limiting the size of the mixed cores for smaller masses.

In Fig.~\ref{figcompareALL} we show a comparison of the mixed core sizes obtained without any cut, with the square cut according to Eq.~(\ref{eqcut2}) and with the tanh cut according to Eq.~(\ref{eqcuttanh}). For reference, results obtained with MLT and the 1- and 3-equation models are shown in the same figure. As discussed above, the square cut is more restrictive than the tanh cut at masses around and above $2\,M_\odot$. Only at a mass of $8\,M_\odot$ the square cut does not restrict the convective core size any more. In contrast, the tanh cut restricts the core size only marginally, already at $2\,M_\odot$. For lower masses below about $4\,M_\odot$, the results of the square cut are in good agreement with the 3-equation model.

\end{document}